\documentclass{article}
\usepackage{arxiv}
\usepackage{times}
\usepackage{epsfig}
\usepackage{graphicx}
\usepackage{amsmath}
\usepackage{amssymb}
\usepackage{adjustbox}
\usepackage{amsfonts}       %
\usepackage[square,sort,comma,numbers]{natbib}
\usepackage[utf8]{inputenc} %
\usepackage[T1]{fontenc}    %
\usepackage{url}
\usepackage{booktabs} %
\usepackage{framed} 
\usepackage{float}
\usepackage{enumitem}
\usepackage{capt-of}
\usepackage{subfig}
\usepackage{booktabs}       %
\usepackage{nicefrac}       %
\usepackage{microtype}      %
\usepackage{xcolor}
\usepackage{xparse}

\usepackage{threeparttable}
\usepackage{tabularx}

\usepackage[title]{appendix}

\begin{document}
\title{Large datasets: A Pyrrhic win for computer vision?}

\author{Vinay Uday Prabhu\thanks{Equal contributions} \\
UnifyID AI Labs\\
Redwood City\\
{\tt\small vinay@unify.id}
\And
Abeba Birhane\footnotemark[1] \\
School of Computer Science, UCD, Ireland \\ Lero - The Irish Software Research Centre\\
{\tt\small abeba.birhane@ucdconnect.ie}
}

\maketitle

\begin{abstract}
In this paper we investigate problematic practices and consequences of large scale vision datasets. We examine broad issues such as the question of consent and justice as well as specific concerns such as the inclusion of verifiably pornographic images in datasets. Taking the ImageNet-ILSVRC-2012 dataset as an example, we perform a cross-sectional model-based quantitative census covering factors such as age, gender, NSFW content scoring, class-wise accuracy, human-cardinality-analysis, and the semanticity of the image class information in order to statistically investigate the extent and subtleties of ethical transgressions. We then use the census to help hand-curate a look-up-table of images in the ImageNet-ILSVRC-2012 dataset that fall into the categories of verifiably pornographic: shot in a non-consensual setting (up-skirt), beach voyeuristic, and exposed private parts. We survey the landscape of harm and threats both society broadly and individuals face due to uncritical and ill-considered dataset curation practices. We then propose possible courses of correction and critique the pros and cons of these. We have duly open-sourced all of the code and the census meta-datasets generated in this endeavor for the computer vision community to build on. 
By unveiling the severity of the threats, our hope is to motivate the constitution of mandatory Institutional Review Boards (IRB) for large scale dataset curation processes. 
\end{abstract}
\section{Introduction}
Born from World War II and the haunting and despicable practices of Nazi era experimentation \cite{Nazianat71:online} the \textit{1947 Nuremberg code}~\cite{weindling2001origins_informed_consent} and the subsequent \textit{1964 Helsinki declaration}~\cite{Timetodi89:online_Nature}, helped establish the doctrine of \textbf{Informed Consent} which builds on the fundamental notions of human dignity and agency to control dissemination of information about oneself. This has shepherded data collection endeavors in the medical and psychological sciences concerning human subjects, including photographic data \cite{naidoo2009informed, blain2002informed}, for the past several decades. A less stringent version of informed consent, \textit{broad consent}, proposed in 45 CFR 46.116(d) of the \textit{Revised Common Rule}~\cite{eCFR_Cod99_online_common_rule}, has been recently introduced that still affords the basic safeguards towards protecting one's identity in large scale databases. However, in the age of \textit{Big Data}, the fundamentals of informed consent, privacy, or agency of the individual have gradually been eroded. Institutions, academia, and industry alike, amass millions of images of people without consent and often for unstated purposes under the guise of anonymization. These claims are misleading given there is weak anonymity and privacy in aggregate data in general \cite{narayanan2008robust_de-anonymization} and more crucially, images of faces are not the type of data that can be aggregated. As can be seen in Table~\ref{tab:datasets}, several tens of millions of images of people are found in peer-reviewed literature. These images are obtained without consent or awareness of the individuals or IRB approval for collection. In \textit{Section 5-B} of~\cite{torralba200880_tiny}, for instance, the authors nonchalantly state ``\textit{As many images on the web contain pictures of people, a large fraction (23\%) of the 79 million images in our dataset have people in them}''. With this background, we now focus on one of the most celebrated and canonical large scale image datasets: the \textit{ImageNet} dataset. From the questionable ways images were sourced, to troublesome labeling of people in images, to the downstream effects of training AI models using such images, ImageNet and large scale vision datasets (LSVD) in general \textit{constitute a Pyrrhic win} for computer vision. We argue, this win has come at the expense of harm to \textit{minoritized groups} and further aided the gradual erosion of privacy, consent, and agency of both the individual and the collective.

\subsection{ImageNet: A brief overview}

\begin{table*}[ht]
\centering
\begin{threeparttable}
\caption{Large scale image datasets containing people's images}
\begin{tabular}{|l|l|l|l|}
\hline
Dataset &
  \begin{tabular}[c]{@{}l@{}}Number of images\\ (in millions)\end{tabular} &
  \begin{tabular}[c]{@{}l@{}}Number of \\ categories\\ (in thousands)\end{tabular} &
  \begin{tabular}[c]{@{}l@{}}Number of \\ consensual\\ images\end{tabular} \\ \hline
JFT-300M (\cite{hinton2015distilling_jft})     & 300+    &  18                 & 0 \\ \hline
Open Images (\cite{kuznetsova2018open})      & 9    &  20                 & 0 \\ \hline
Tiny-Images (\cite{torralba200880_tiny}) & 79 & 76 & 0 \\ \hline
Tencent-ML (\cite{wu2019tencent})  & 18     & 11                 & 0 \\ \hline
ImageNet-(21k,11k\tnote{1},1k) (\cite{russakovsky2015imagenet})  & (14, 12, 1) & (22, 11, 1)                 & 0 \\ \hline
Places (\cite{zhou2017places})      & 11 & 0.4                   & 0 \\ \hline
\end{tabular}%
\begin{tablenotes}
\item [1] See \url{https://modelzoo.co/model/resnet-mxnet}
\end{tablenotes}
\label{tab:datasets}
\end{threeparttable}
\end{table*}
The emergence of the ImageNet dataset \cite{deng2009imagenet} is widely considered a pivotal moment\footnote{``The data that transformed AI research—and possibly the world'': \url{https://bit.ly/2VRxx3L}} in the Deep Learning revolution that transformed Computer Vision (CV), and Artificial Intelligence (AI) in general. 
Prior to ImageNet, computer vision and image processing researchers trained image classification models on small datasets such as CalTech101 (9k images), PASCAL-VOC (30k images), LabelMe (37k images), and the SUN (131k images) dataset (see slide-37 in \cite{imagenet30:online}). ImageNet, with over 14 million images spread across 21,841 synsets, replete with 1,034,908 bounding box annotations, brought in an aspect of scale that was previously missing. A subset of 1.2 million images across 1000 classes was carved out from this dataset to form the ImageNet-1k dataset (popularly called \texttt{ILSVRC-2012}) which formed the basis for the \textit{Task-1: classification} challenge in the ImageNet Large Scale Visual Recognition Challenge (ILSVRC). This soon became widely touted as the \textit{Computer Vision Olympics}\footnote{ https://engineering.missouri.edu/2014/01/team-takes-top-rankings-in-computer-vision-olympics/}. The vastness of this dataset allowed a Convolutional Neural Network (CNN) with 60 million parameters \cite{krizhevsky2012imagenet} trained by the \textit{SuperVision} team from University of Toronto to usher in the rebirth of the CNN-era (see \cite{alom2018history_alexnet}), which is now widely dubbed the \textit{AlexNet moment} in AI.

Although ImageNet was created over a decade ago, it remains one of the most influential and powerful image databases available today. Its power and magnitude is matched by its unprecedented societal impact. Although an \textit{a posteriori} audit might seem redundant a decade after its creation, ImageNet's continued significance and the culture it has fostered for other large scale datasets warrants an ongoing critical dialogue.

The rest of this paper is structured as follows. In section 2, we cover related work that has explored the ethical dimensions that arise with LSVD. In section 3, we describe the landscape of both the immediate and long term threats individuals and society as a whole encounter due to ill-considered LSVD curation. In Section 4, we propose a set of solutions which might assuage some of the concerns raised in section 3. In Section 5, we present a template quantitative auditing procedure using the ILSVRC2012 dataset as an example and describe the data assets we have curated for the computer vision community to build on. We conclude with broad reflections on LSVDs, society, ethics, and justice.

\section{Background and related work}
The very declaration of a taxonomy brings some things into existence while rendering others invisible \cite{bowker2000sorting}. A gender classification system that conforms to essentialist binaries, for example, operationalizes gender in a cis-centric way resulting in exclusion of non-binary and transgender people \cite{keyes2018misgendering}. Categories simplify and freeze nuanced and complex narratives, obscuring political and moral reasoning behind a category. Over time, messy and contingent histories hidden behind a category are forgotten and trivialized \cite{star2007enacting}. With the adoption of taxonomy sources, image datasets inherit seemingly invisible yet profoundly consequential shortcomings. The dataset creation process, its implication for ML systems, and subsequently, the societal impact of these systems has attracted a substantial body of critique. We categorize such body of work into two groups that complement one another. While the first group can be seen as concerned with the broad downstream effects, the other concentrates mainly on the dataset creation process itself. 
\subsection{Broad critiques}
The absence of critical engagement with canonical datasets disproportionately negatively impacts women, racial and ethnic minorities, and vulnerable individuals and communities at the margins of society~\cite{birhane2019algorithmic}. For example, image search results both exaggerate stereotypes and systematically under-represent women in search results for occupations~\cite{kay2015unequal}; object detection systems designed to detect pedestrians display higher error rates for recognition of demographic groups with dark skin tones~\cite{wilson2019predictive}; and gender classification systems show disparities in image classification accuracy where lighter-skin males are classified with the highest accuracy while darker-skin females suffer the most misclassification~\cite{buolamwini2018gender}. Gender classification systems that lean on binary and cis-genderist constructs operationalize gender in a trans-exclusive way resulting in tangible harm to trans people~\citep{keyes2018misgendering, scheuerman2019computers}. With a persistent trend where minoritized and vulnerable individuals and communities often disproportionately suffer the negative outcomes of ML systems, D'Ignazio and Klein \cite{d2020data} have called for a shift in rethinking ethics not just as a fairness metric to mitigate the narrow concept of bias but as a practice that results in justice for the most negatively impacted. Similarly, Kasy and Abebe \cite{kasy2020fairness} contend that perspectives that acknowledge existing inequality and aim to redistribute power are pertinent as opposed to fairness-based perspectives. Such understanding of \textit{ethics as justice} then requires a focus beyond `bias' and fairnesss' in LSVDs and requires questioning of how images are sourced, labelled, and what it means for models to be trained on them. 
One of the most thorough investigations in this regard can be found in \cite{Excavati19:online}. In this recent work, Crawford and Paglen present an in-depth critical examination of ImageNet including the dark and troubling results of classifying people as if they are objects. Offensive and derogatory labels that perpetuate historical and current prejudices are assigned to people’s actual images. The authors emphasise that not only are images that were scraped across the web appropriated as data for computer vision tasks, but also the very act of assigning labels to people based on physical features raises fundamental concerns around reviving long-discredited pseudo-scientific ideologies of physiognomy~\cite{y2017physiognomy}. 
\subsection{Critiques of the curation phase}
Within the dataset creation process, \textit{taxonomy sources} pass on their limitations and underlying assumptions that are problematic. The adoption of underlying structures presents a challenge where --- without critical examination of the architecture --- ethically dubious taxonomies are inherited. This has been one of the main challenges for ImageNet given that the dataset is built on the backbone of WordNet's structure. Acknowledging some of the problems, the authors from the ImageNet team did recently attempt to address \cite{yang2020towards_facct} the stagnant concept vocabulary of WordNet. They admitted that only 158 out of the 2,832 existing synsets should remain in the person sub-tree\footnote{In order to prune all the nodes. They also took into account the \textit{imageability} of the synsets and the skewed representation in the images pertaining to the \textit{Image retrieval} phase}. Nonetheless, some serious problems remain untouched. This motivates us to address in greater depth the overbearing presence of the \textit{WordNet effect} on image datasets. 

\subsection{The WordNet Effect}
\begin{figure}[ht!]
\centering
\subfloat[Class-wise counts of the offensive classes]{
  \includegraphics[width=0.45\textwidth]{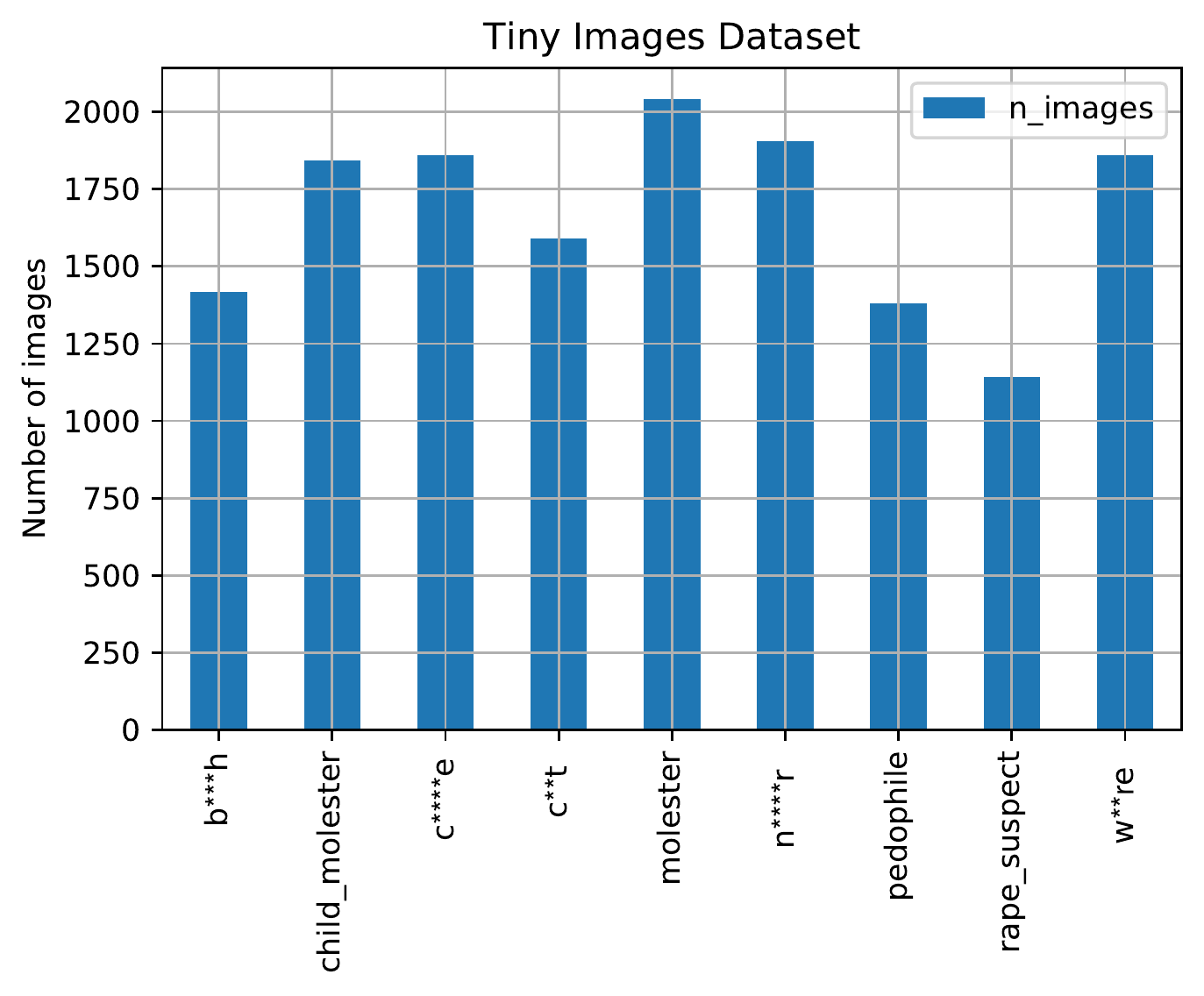}
}
\subfloat[Samples from the class labelled \texttt{n****r}]{
  \includegraphics[width=0.375\textwidth]{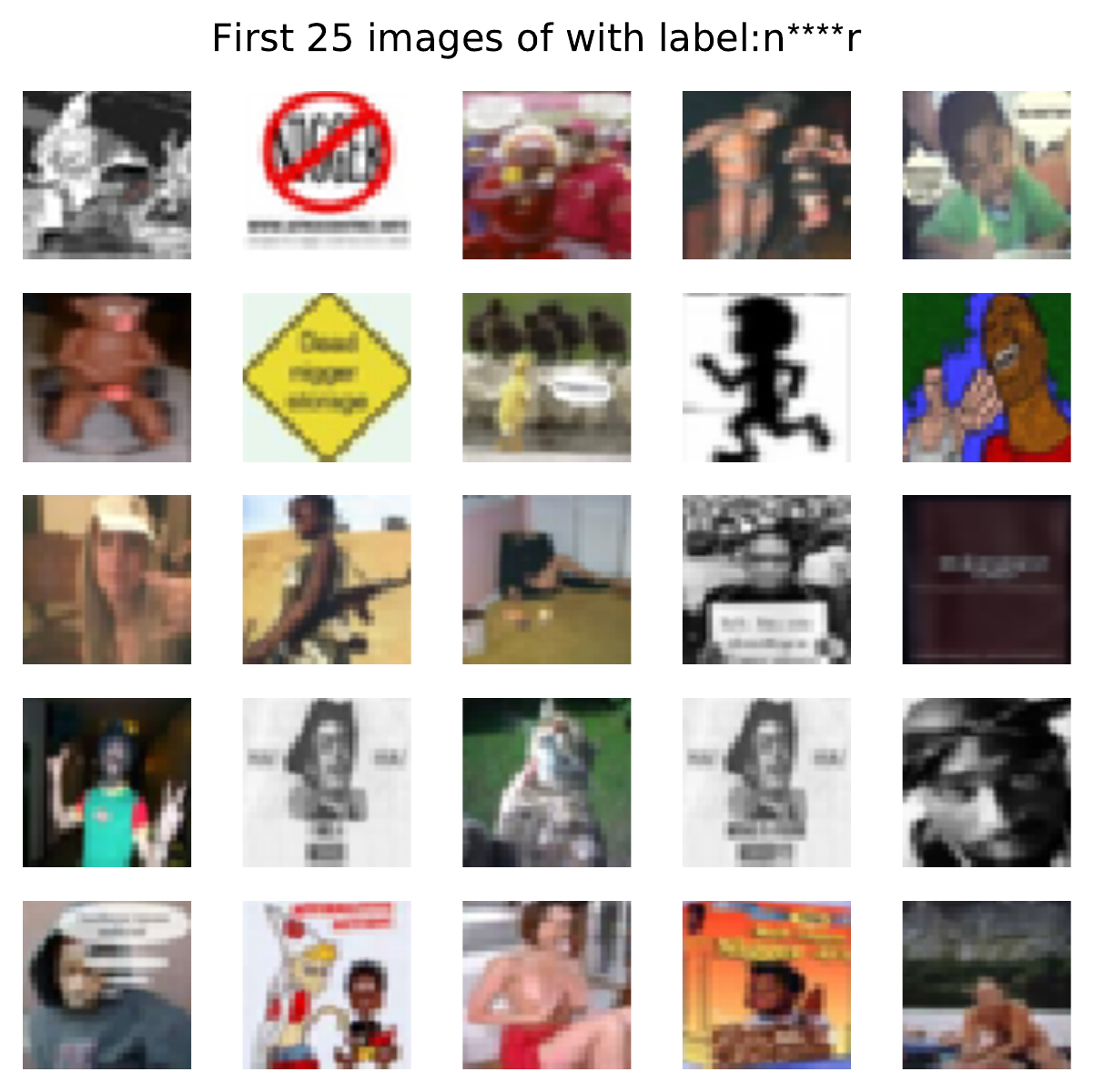}
}
\caption{Results from the \textit{80 Million Tiny Images} dataset exemplifying the toxicities of it's label space}
\label{fig:tiny_images}
\end{figure}

ImageNet is not the only large scale vision dataset that has inherited the shortcomings of the WordNet taxonomy. The \textit{80 million Tiny Images} dataset \cite{torralba200880_tiny} which grandfathered the CIFAR-10/100 datasets and the Tencent ML-images dataset \cite{wu2019tencent} also used the same path. Unlike ImageNet, these datasets have never been audited\footnote{In response to the mainstream media covering a pre-print of this work, we were informed that the curators of the dataset have withdrawn the dataset with a note accessible here: \url{https://groups.csail.mit.edu/vision/TinyImages/}} or scrutinized and some of the sordid results from inclusion of \textit{ethnophaulisms} in Tiny-Images dataset's label space are displayed in Figure \ref{fig:tiny_images}. The figure demonstrates both the number of images in a subset of the \textit{offensive} classes (sub-figure(a)) and the exemplar images (sub-figure(b)) that show the images in the noun-class labelled \texttt{n****r\footnote{Due to its offensiveness, we have censored this word (and other words throughout the paper), however, it remains uncensored on the website at the time of writing.}}, a fact that serves as a stark reminder that a great deal of work remains to be done by the ML community at large.

Similarly, we found at least 315 classes\footnote{See \url{https://bit.ly/30DybmF}} of the potentially 1593 classes deemed to be \textit{non-imageable} by the ImageNet curators in \cite{yang2020towards_facct} still retained in the Tencent-ML-Images dataset that includes image classes such as \texttt{[transvestite, bad person,  fornicatress, orphan, mamma's boy, and enchantress]}.

Finally, the \textit{labeling and validation} of the curation process also present ethical challenges. Recent work such as~\cite{gray2019ghost} has explored the intentionally hidden labour, which they have termed as \textit{Ghost Work}, behind such tasks. Image labeling and validation requires the use of crowd-sourced platforms such as MTurk, often contributing to the exploitation of underpaid and undervalued \textit{gig workers}. Within the topic of image labeling but with a different dimension and focus, recent work such as \cite{imagenet_mit} and \cite{2006.07159_arewedone} has focused on the shortcomings of human-annotation procedures used during the ImageNet dataset curation. These shortcomings, the authors point out, include single label per-image procedure that causes problems given that real-world images often contain multiple objects, and inaccuracies due to ``overly restrictive label proposals''. 

\section{The threat landscape}
\label{sec: threats}
In this section, we survey the landscape of harm and threats, both immediate and long term, that emerge with dataset curation practices in the absence of careful ethical considerations and anticipation for negative societal consequences. Our goal here is to bring awareness to the ML and AI community regarding the severity of the threats and to motivate a sense of urgency to act on them. We hope this will result in practices such as the mandatory constitution of Institutional Review Boards (IRB) for large scale dataset curation processes.

\subsection{The rise of reverse image search engines, loss  of privacy, and the blackmailing threat}
Large image datasets, when built without careful consideration of societal implications, pose a threat to the welfare and well-being of individuals. Most often, vulnerable people and marginalised populations pay a disproportionately high price. Reverse image search engines that allow face search such as \cite{Facesear28:online} have gotten remarkably and worryingly efficient in the past year. For a small fee, anyone can use their portal or their API\footnote{Please refer to the supplementary material in Appendix \ref{appendix:privacy_loss} for the screenshots} to run an automated process to uncover the ``real-world'' identities of the \textit{humans of ImageNet} dataset. For example, in societies where sex work is socially condemned or legally criminalized, re-identification of a sex worker through image search, for example, bears a real danger for the individual victim. %
Harmful discourse such as \textit{revenge porn}, are part of a broader continuum of image-based sexual abuse \cite{mcglynn2017beyond}. To further emphasize this specific point, many of the images in classes such as \texttt{maillot}, \texttt{brassiere}, and \texttt{bikini} contain images of beach voyeurism and other non-consensual cases of digital image gathering (covered in detail in Section \ref{sec:quant}). We were able to (unfortunately) easily map the victims, most of whom are women, in the pictures to ``real-world'' identities of people belonging to a myriad of backgrounds including teachers, medical professionals, and academic professors using reverse image search engines such as \cite{Facesear8:online}. Paying heed to the possibility of the \textit{Streisand effect}\footnote{The Streisand effect \textit{``is a social phenomenon that occurs when an attempt to hide, remove, or censor information has the unintended consequence of further publicizing that information, often via the Internet''} \cite{Streisan38:online}
}, we took the decision not to divulge any further quantitative or qualitative details on the extent or the location of such images in the dataset besides alerting the curators of the dataset(s) and making a passionate plea to the community not to underestimate the severity of this particular threat vector. 
\subsection{The emergence of even larger and more opaque datasets} The attempt to build computer vision has been gradual and can be traced as far back as 1966 to Papert's \textit{The Summer Vision Project}~\citep{papert1966summer}, if not earlier. However, ImageNet, with its vast amounts of data, has not only erected a canonical landmark in the history of AI, it has also paved the way for even bigger, more powerful, and suspiciously opaque datasets. The lack of scrutiny of the ImageNet dataset by the wider computer vision community has only served to embolden institutions, both academic and commercial, to build far bigger datasets without scrutiny (see Table \ref{tab:datasets}). Various highly cited and celebrated papers in recent years~\cite{hinton2015distilling_jft,chollet2017xception_jft,brock2018large_jft,sun2017revisiting_jft}, for example, have used the \textit{unspoken unicorn} amongst large scale vision datasets, that is, the JFT-300M dataset [?]\footnote{We have decided to purposefully leave the '?' in place and plan to revisit it only after the dataset's creator(s) publish the details of it's curation}. This dataset is inscrutable and operates in the dark, to the extent that there has not even been official communication as to what \textit{JFT-300M} stands for. All that the ML community knows is it purportedly boasts more than 300M images spread across 18k categories. The open source variant(s) of this, the \textit{Open Images V4-5-6}~\cite{kuznetsova2018open} contains a subset of 30.1M images covering 20k categories (and also has an extension dataset with 478k crowd-sourced images across more than 6000 categories).
While parsing through some of the images, we found \textbf{verifiably}\footnote{See \url{https://bit.ly/2y1sC7i}. We performed verification with the uploader of the image via the Flickr link shared.} non-consensual images of children that were siphoned off of \textit{flickr} hinting towards the prevalence of similar issues for JFT-300M from which this was sourced. Besides the other large datasets in Table \ref{tab:datasets}, we have cases such as the \textit{CelebA-HQ} dataset, which is actually a \textit{heavily processed} dataset whose grey-box curation process only appears in Appendix-C of \cite{karras2017progressive} where no clarification is provided on this \textit{"frequency based visual quality metric"} used to sort the images based on \textit{quality}. Benchmarking any downstream algorithm of such an opaque, biased and (semi-)synthetic dataset will only result in controversial scenarios such as \cite{pulse}, where the authors had to hurriedly incorporate addendums admitting biased results.
Hence, it is important to reemphasize that the existence and use of such datasets bear direct and indirect impact on people, given that decision making on social outcomes increasingly leans on ubiquitously integrated AI systems trained and validated on such datasets. Yet, despite such profound consequences, critical questions such as where the data comes from or whether the images were obtained consensually are hardly considered part of the LSVD curation process. 

The more nuanced and perhaps indirect impact of ImageNet is the \textit{culture} that it has cultivated within the broader AI community; a culture where the appropriation of images of real people as raw material free for the taking has come be to perceived as \textit{the norm}. Such norm and lack of scrutiny has played a role towards the creation of monstrous and secretive datasets without much resistance, prompting further questions such as `what other secretive datasets currently exist hidden and guarded under the guise of proprietary assets?'. 
Current work that has sprung out of secretive datasets, such as Clearview AI~\cite{clearview_nyt}
\footnote{Clearview AI is a US based privately owned technology company that provides facial recognition services to various customers including North American law enforcement agencies. With more than 3 billion photos scraped from the web, the company operated in the dark until its services to law enforcement was reported in late 2019},
points to a deeply worrying and insidious threat not only to vulnerable groups but also to the very meaning of privacy as we know it \cite{ACLUsues41:online_khari}.
\subsection{The Creative Commons fallacy} In May 2007 the iconic case of \textit{Chang versus Virgin mobile: The school girl, the billboard, and virgin} \cite{ChangvVirgin59:online} unraveled in front of the world, leading to widespread debate on the uneasy relationship between personal privacy, consent, and image copyright, initiating a substantial corpus of academic debate (see \cite{corbett2009creative, corbett2011creative, hietanen2011creative, carroll2011school_creative}). A Creative Commons license addresses only copyright issues -- not privacy rights or consent to use images for training. Yet, many of the efforts beyond ImageNet, including the Open Images dataset \cite{kuznetsova2018open}, have been built on top of the \textit{Creative commons} loophole that large scale dataset curation agencies interpret as a \textit{free for all, consent-included} green flag. This, we argue, is fundamentally fallacious as is evinced in the views presented in \cite{UseandFa55:CCBY} by the Creative commons organization that reads: 
\textit{``CC licenses were designed to address a specific constraint, which they do very well: unlocking restrictive copyright. But copyright is not a good tool to protect individual privacy, to address research ethics in AI development, or to regulate the use of surveillance tools employed online.''}. Datasets culpable of this \textit{CC-BY heist} such as \textit{MegaFace} and \textit{IBM's Diversity in Faces} have now been deleted in response to the investigations (see \cite{2001.03071_ibm_ms} for a survey) lending further support to the Creative Commons fallacy. 
\subsection{Blood diamond effect in models trained on this dataset} Akin to the \textit{ivory carving-illegal poaching}, and \textit{diamond jewelry art-blood diamond} nexuses, we posit that there is a similar moral conundrum at play here that effects all downstream applications entailing models trained using a \textit{tainted} dataset. Often, these transgressions may be rather subtle. In this regard, we pick an examplar field of application that on the surface appears to be a low risk application area: \textit{Neural generative art}. Neural generative art created using tools such as BigGAN \cite{brock2018large_jft} and Art-breeder \cite{GANbreeder} that in turn use pre-trained deep-learning models trained on ethically dubious datasets, bear the downstream burden\footnote{Please refer to the appendix ( Section \ref{appendix:blood_diamond}) where we demonstrate one such real-world experiment entailing unethically generated neural art replete with responses obtained from human critiques as to what they felt about the imagery being displayed.} of the problematic residues from non-consensual image siphoning, thus running afoul of the Wittgensteinian edict of \textit{ethics and aesthetics being one and the same.} \cite{eaglestone2004one}. We also note that there is a privacy-leakage facet to this \textit{downstream burden}. In the context of face recognition,  works such as \cite{song2017machine_toomuch} have demonstrated that CNNs with high predictive power unwittingly accommodate accurate extraction of subsets of the facial images that they were trained on, thus abetting dataset leakage\footnote{We'd like to especially highlight the \textit{megapixel.cc} project \cite{MegaPixels.cc} for the ground-breaking work on datasets to train such facial recognition systems }. 
\subsection{Perpetuation of unjust and harmful stereotypes} Finally, zooming out and taking a broad perspective allows us to see that the very practice of embarking on a classification, taxonomization, and labeling task endows the classifier with the power to decide what is a legitimate, normal, or correct way of being, acting, and behaving in the social world \cite{bowker2000sorting}. For any given society, what comes to be perceived as \textit{normal} or \textit{acceptable} is often dictated by dominant ideologies. Systems of classification, which operate within a power asymmetrical social hierarchy, necessarily embed and amplify historical and cultural prejudices, injustices, and biases \cite{star2007enacting}. In western societies, ``desirable'', ``positive'', and ``normal'' characteristics and ways of being are constructed and maintained in ways that align with the dominant narrative, giving advantage to those that fit the status quo. Groups and individuals on the margins, on the other hand, are often perceived as the ``outlier'' and the ``deviant''. Image classification and labelling practices, without the necessary precautions and awareness of these problematic histories, pick up these stereotypes and prejudices and perpetuate them \cite{o2016weapons, noble2018algorithms, eubanks2018automating}. AI systems trained on such data amplify and normalize these stereotypes, inflicting unprecedented harm on those that are already on the margins of society. While the ImageNet team did initiate strong efforts towards course-correction \cite{yang2020towards_facct}, the Tiny Images dataset still contains harmful slurs and offensive labels. And worse, we remain in the dark regarding the secretive and opaque LSVDs. 
\begin{figure*}
\raggedright
\begin{framed}
\begin{center}{\LARGE {\bf Dataset audit card - ImageNet}}\end{center} 
\vspace{-0.15in}

\begin{minipage}{.475\textwidth}
{\bf Census audit statistics} 
\begin{itemize}[leftmargin=*]
\item  $83436$ images with $101070-132201$ persons (Models: DEX (\cite{rothe2018deep_dex}), {InsightFace} (\cite{insight-face}))
\item Mean-age (male): 33.24 (Female):25.58 ( RetinaFace \cite{deng2019retinaface}, ArcFace \cite{deng2018arcface})
\item Confirmed misogynistic images: 62. Number of classes with infants: 30
\item (${\mu _c^{(A)}}$ and ${\sigma_c^{(A)}}$: Mean and standard-deviation of the gender-estimate of images in class $c$ estimated by  algorithm $(A)$.)
\end{itemize}

\end{minipage}
\vspace{-0.15in}
\raggedleft
\begin{minipage}{.475\textwidth}
{\bf Metrics:} 
Class-level mean count ($\eta _c^{(A)}$), mean gender  \ \ \ \    skewness ($\xi _c^{(A)}$) and mean-age ($\alpha_c^{(A)}$): \begin{align*} 
\eta _c^{(A)} = \frac{1}{{{N_c}}}\sum\limits_{i = 1}^{{N_c}} {I[{\phi _i}]} , \alpha _c^{(A)} = \frac{1}{{{N_c}}}\sum\limits_{i = 1}^{{N_c}} {I[{\phi _i}]a_i^{(A)}} \;and\;\\
\xi _c^{(A)} = \frac{1}{{{N_c}}}{\sum\limits_{i = 1}^{{N_c}} {I[{\phi _i}]\left( {\frac{{g_i^{(A)} - \mu _c^{(A)}}}{{\sigma _c^{(A)}}}} \right)} ^3} 
\label{eq:metrics_scatter}
\end{align*} 
\ \ \  \ \ \ $
{\phi _i} = 
     \begin{cases}
       \text{1} &\text{if face present }\\
       \text{0} &\text{otherwise} \\
     \end{cases}
$
 in $i^{th}$ image.
\end{minipage}
\begin{figure}[H]
  \centering
  \includegraphics[width=0.95\textwidth, height=1.95in]{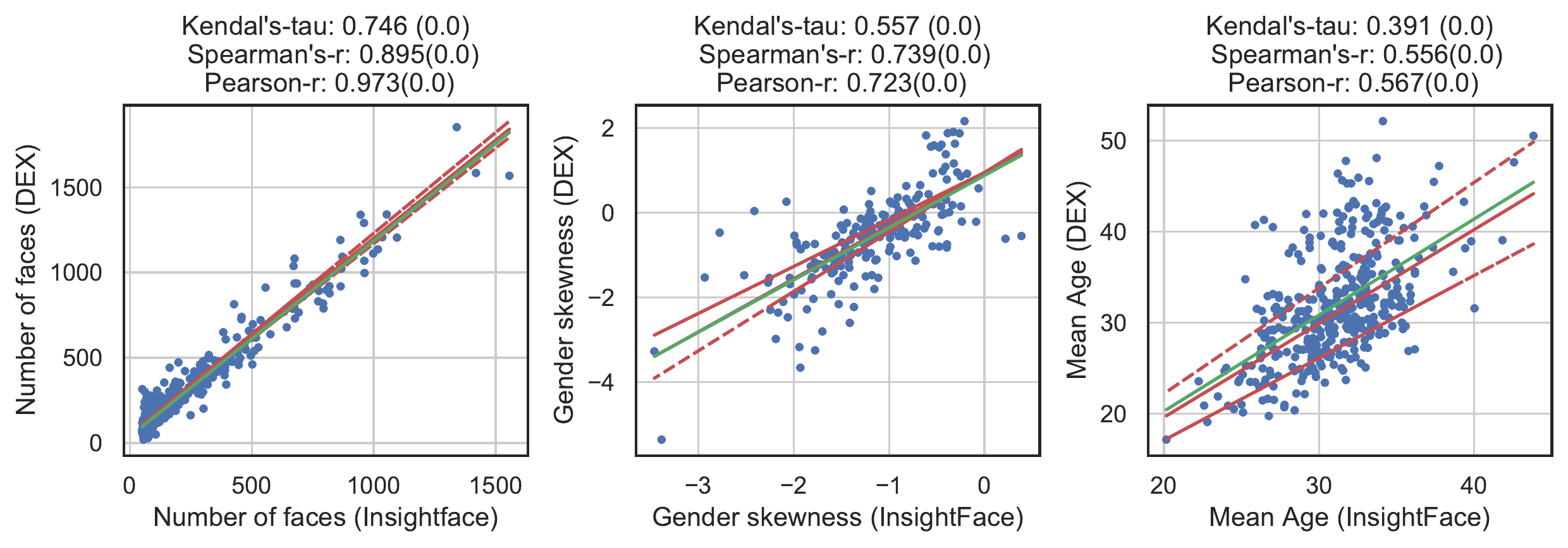}
  \caption{Class-wise cross-categorical scatter-plots across the cardinality, age and gender scores}
  \label{fig:cag_card}
\end{figure}
\vspace{-0.35in}
\begin{figure}[H]
  \centering
  \includegraphics[width=0.9\textwidth, height=1.95in]{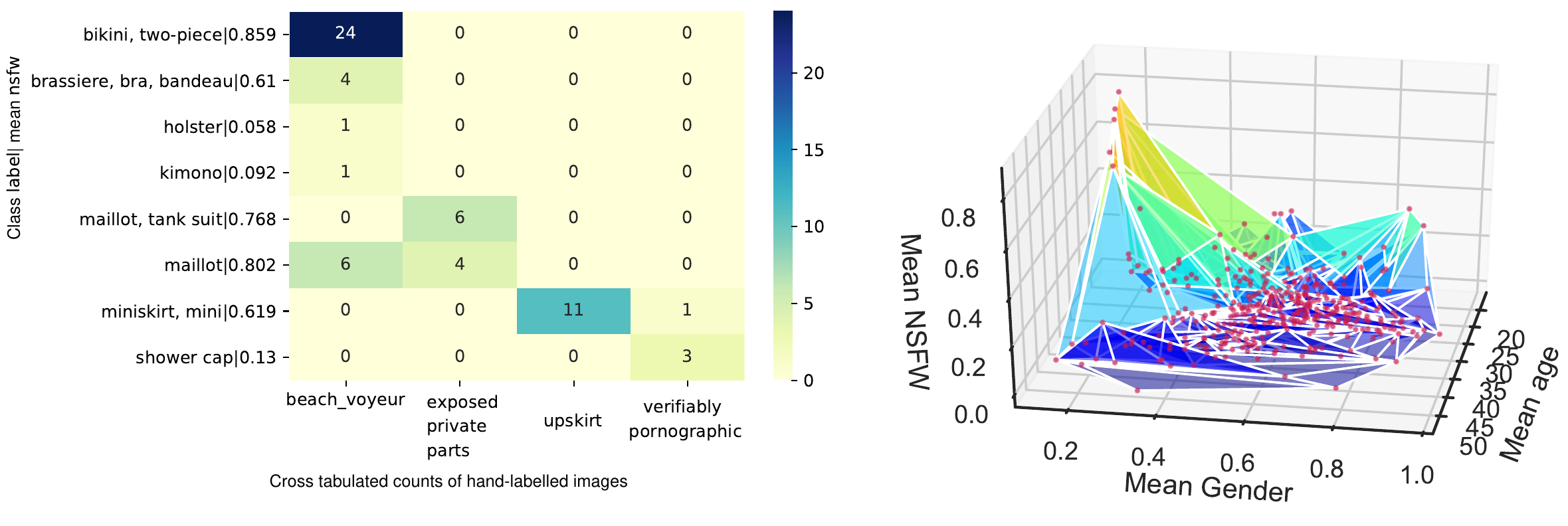}
  \caption{Statistics and locationing of the hand-labelled images}
  \label{fig:agn_card}
\end{figure}
\vspace{-0.35in}
\begin{figure}[H]
  \centering
  \includegraphics[width=0.9\textwidth, height=1.75in]{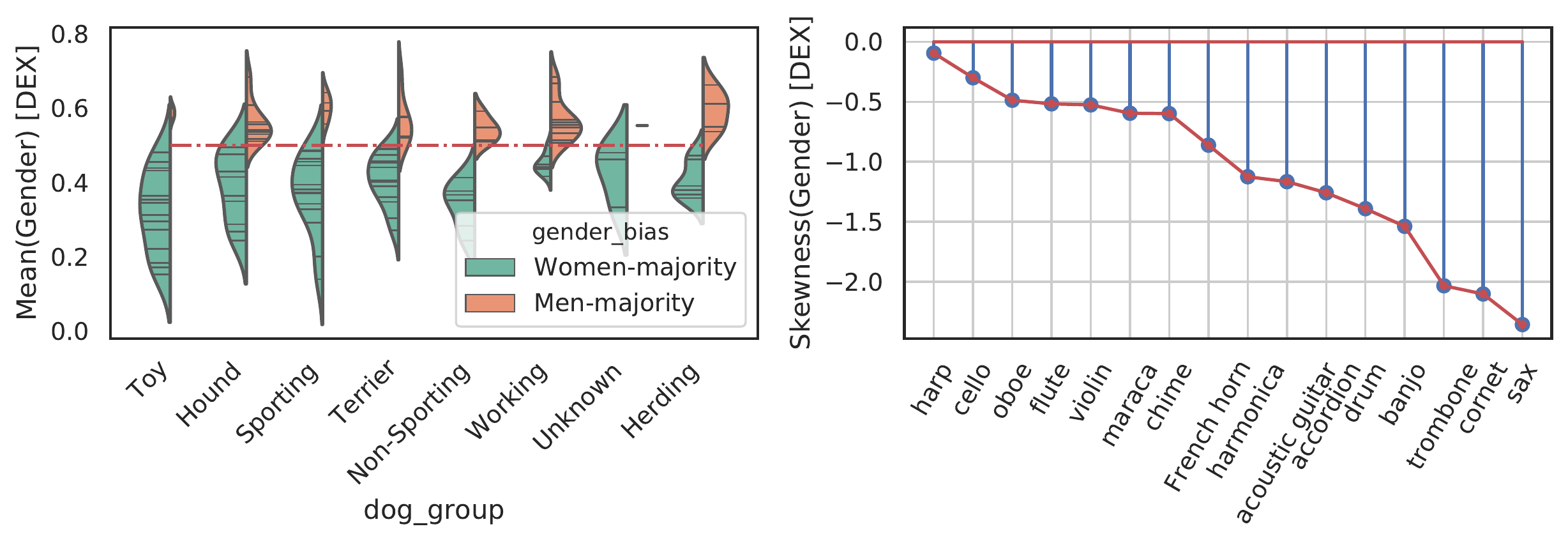}
  \vspace*{-2mm}
  \caption{Known \textit{human co-occurrence} based gender-bias analysis}
  \label{fig:co_occur_card}
\end{figure}
\end{framed}
\caption{Dataset audit card for the ImageNet dataset}
\label{fig:audit_card}
\end{figure*}

\section {Candidate solutions: The path ahead}
Decades of work within the fields of Science and Technology Studies (STS) and the Social Sciences show that there is no single straightforward solution to most of the wider social and ethical challenges that we have discussed \cite{suchman2007human,benjamin2019race,d2020data}. These challenges are deeply rooted in social and cultural structures and form part of the fundamental social fabric. Feeding AI systems on the world's beauty, ugliness, and cruelty, but expecting it to reflect only the beauty is a fantasy \cite{benjamin2019race}. These challenges and tensions will exist as long as humanity continues to operate. Given the breadth of the challenges that we have faced, any attempt for a quick fix risks concealing the problem and providing a false sense of solution. The idea of a complete removal of biases, for example, might in reality be simply hiding them out of sight \cite{gonen2019lipstick}. Furthermore, many of the challenges (bias, discrimination, injustice) vary with context, history, and place, and are concepts that continually shift and change constituting a moving target \cite{birhane2019algorithmic}. The pursuit of panacea in this context, therefore, is not only unattainable but also misguided. Having said that, there are remedies that can be applied to overcome the specific harms that we have discussed in this paper, which eventually potentially play constituent roles in improving the wider and bigger social and structural issues in the long run. 

\subsection{Remove, replace, and open strategy}
In \cite{yang2020towards_facct}, the authors concluded that within the \textit{person sub-tree} of the ImageNet dataset, 1593 of the 2832 people categories were \textit{potentially offensive} labels and planned to \textit{"remove all of these from ImageNet."}. We strongly advocate a similar path for the offensive noun classes in the Tiny Images dataset that we have identified in section 2.1, as well as images that fall into the categories of verifiably\footnote{We use the term \textit{verifiably} to denote only those NSFW images that were hand-annotated by the volunteers indicating that they also contained the textual context that was of pornographic phraseology. We have an example grid of these images in the Appendix.} pornographic, shot in a non-consensual setting (up-skirt), beach voyeuristic, and exposed genitalia in the ImageNet-ILSVRC-2012 dataset. In cases where the image category is retained but the images are not, the option of \textit{replacement with consensually shot financially compensated images} arises. It is possible that some of the people in these images might come forward to consent and contribute their images in exchange for fair financial compensation, credit, or out of sheer altruism \cite{brown2019people}. We re-emphasize that our consternation focuses on the non-consensual aspect of the images and not on the category-class and the ensuing content of the images in it. This solution, however, brings forth further questions: does this make image datasets accessible only to those who can afford it? Will we end up with pool of images with a predominantly financially disadvantaged participants? 

Science is self-correcting so long as it is accessible and open to critical engagement. We have tried to engage critically and map actionable ways forward given what we know of these LSVDs. The secretive and opaque LSVDs, however, thread a dangerous territory, given that they directly or indirectly impact society but remain hidden and inaccessible. Although the net benefit of the open science movement remains controversial, we strongly contend that making LSVDs open and accessible allows audits of these datasets, which is a first step towards a responsible scientific endeavour. 

\subsection{Automated downstream removal from reverse search engines that allow for image deletion requests}
We found that some of the reverse image search engines do allow for \textit{users to remove particular image from our [sic] index} via their \textit{"Report abuse"} portals\footnote{See \url{https://pimeyes.com/en/faq/remove-from-database}}. This allows for dataset auditors to enlist images found in their dataset(s) containing identifiable individuals and direct them towards a guided image removal process from the reverse image search engine(s), in order to mitigate some aspects of immediate harm.

\subsection{Differentially private obfuscation of the faces} 
This path entails harnessing techniques such as DP-Blur \cite{fan2018image} with quantifiable privacy guarantees to obfuscate the identity of the humans in the image. The \textit{Inclusive images challenge} \cite{shankar2017no_diversity}, for example, already incorporated blurring during dataset curation\footnote{\url{https://www.kaggle.com/c/inclusive-images-challenge} } and addressed the downstream effects surrounding change in predictive power of the models trained on the blurred versions of the dataset curated. We believe that replication of this template that also clearly included avenues for recourse in case of an erroneously non-blurred image being sighted by a researcher will be a step in the right direction for the community at large.
\subsection{Synthetic-to-real and Dataset distillation} 
The basic idea here is to utilize (or augment) synthetic images in lieu of real images during model training. 
Approaches include using hand-drawn sketch images (\textit{ImageNet-Sketch}~\citep{wang2019learning}), using GAN generated images \cite{GAN_aug} and techniques such as \textit{Dataset distillation}~\cite{wang2018dataset}, where a dataset or a subset of a dataset is distilled down to a few representative \textit{\textbf{synthetic}} samples. This is a nascent field with some promising results emerging in unsupervised domain adaptation across visual domains \cite{peng2018visda} and universal digit classification \cite{prabhu2019fonts}.
\subsection{Ethics-reinforced filtering during the curation}
The specific ethical transgressions that emerged during our longitudinal analysis of ImageNet could have been prevented if there were explicit instructions provided to the \textit{MTurkers} during the dataset curation phase to enable filtering of these images at the source (See Fig.9 in~\cite{recht2019imagenet_v2} for example). We hope ethics checks become an integral part of the User-Interface deployed during the humans-in-the-loop validation phase for future dataset curation endeavors. 
\subsection{Dataset audit cards} 
As emphasized in Section 4, context is crucial in determining whether a certain dataset ethical or problematic as it provides a vital background information and datasheets are an effective way of providing context. Much along the lines of \textit{model cards} \cite{model_cards} and \textit{datasheet for datasets} \cite{gebru2018datasheets}, we propose dissemination of \textit{dataset audit cards}. This allows large scale image dataset curators to publish the goals, curation procedures, known shortcomings and caveats alongside their dataset dissemination. In Figure \ref{fig:audit_card}, we have curated an example dataset audit card for the ImageNet dataset using the quantitative analyses carried out in Section~\ref{sec:quant}
\section{Quantitative dataset auditing: ImageNet as a template}
\label{sec:quant}
\begin{table*}[ht!]
\centering
\caption{Meta datasets curated during the audit processes}
\begin{tabularx}{\textwidth}{|l|l|X|}
\hline
file\_name &
  shape &
  file\_contents \\ \hline
df\_insightface\_stats.csv &
  (1000, 30) &
  24 classwise statistical parameters obtained by running the \texttt{InsightFace} model (\cite{insight-face}) on the ImageNet dataset \\ \hline
df\_audit\_age\_gender\_dex.csv &
  (1000, 12) &
  11 classwise (ordered by the wordnet-id) statistical parameters obtained from the json files (of the DEX paper) \cite{rothe2018deep_dex} \\ \hline
df\_nsfw.csv &
  (1000, 5) &
  The mean and std of the NSFW scores of the train and val images arranged per-class. (Unnamed: 0: WordNetID of the class) \\ \hline
df\_acc\_classwise\_resnet50.csv &
  (1000, 7) &
  Classwise accuracy metrics (\& the image level preds) obtained by running the ResNet50 model on ImageNet train and Val sets \\ \hline
df\_acc\_classwise\_NasNet\_mobile.csv &
  (1000, 7) &
  Classwise accuracy metrics (\& the image level preds) obtained by running the NasNet model on ImageNet train and Val sets \\ \hline
df\_imagenet\_names\_umap.csv &
  (1000, 5) &
  Dataframe with 2D UMAP embeddings of the Glove vectors of the classes of the ImageNet dataset \\ \hline
df\_census\_imagenet\_61.csv &
  (1000, 61) &
  The MAIN census dataframe covering class-wise metrics across 61 parameters, all of which are explained in df\_census\_columns\_interpretation.csv \\ \hline
df\_census\_columns\_interpretation.csv &
  (61, 2) &
  The interpretations of the 61 metrics of the census dataframe above \\ \hline
df\_hand\_survey.csv &
  (61, 3) &
  Dataframe contaimning the details of the 61 images unearthed via hand survey (Do not pay heed to 61. it is a mere coincidence) \\ \hline
df\_classes\_tiny\_images\_3.csv &
  (75846, 3) &
  Dataframe containing the class\_ind, class\_name (wordnet noun) and n\_images \\ \hline
df\_dog\_analysis.csv &
  (7, 4) &
  Dataframe containing breed, gender\_ratio and survey result from the paper Breed differences in canine aggression' \\ \hline
\end{tabularx}
\label{tab:csv_census}
\end{table*}
 We performed a cross-categorical quantitative analysis of ImageNet to assess the extent of the ethical transgressions and the feasibility of model-annotation based approaches. This resulted in an \textit{ImageNet census}, entailing both image-level as well as class-level analysis across the $57$ different metrics (see supplementary section) covering Count, Age and  Gender (CAG), NSFW-scoring, semanticity of class labels and accuracy of classification using pre-trained models. We have distilled the important revelations of this census as a \textit{dataset audit card} presented in Figure \ref{fig:audit_card}. This audit also entailed a human-in-the-loop based hybrid-approach that the pre-trained-model annotations (along the lines of \cite{dulhanty2019auditing,yang2020towards_facct}) to segment the large dataset into smaller sub-sets and hand-label the smaller subsets to generate two lists covering 62 misogynistic images and 30 image-classes with co-occuring children.
 We used the \texttt{DEX} \cite{rothe2018deep_dex} and the \texttt{InsightFace} \cite{insight-face} pre-trained models\footnote{While harnessing these pre-trained gender classification models, we would like to \textbf{strongly emphasize}  that the specific models and the \textit{problems} that they were intended to solve, when taken in isolation, stand on ethically dubious grounds themselves. In this regard, we strongly concur with previous work such as \cite{ainow_1}  that \textit{gender classification} based on appearance of a person in a digital image is both \textbf{scientifically flawed} and is a technology that bears a high risk of systemic abuse.} to generate the cardinality, gender skewness, and age-distribution results captured in Figure~\ref{fig:cag_card}. This resulted in discovery of \textbf{83,436} images with persons, encompassing \textbf{101,070 to 132,201} individuals, thus constituting $8-10\%$ of the dataset. Further, we munged together gender, age,  class semanticity\footnote{ Obtained using \textit{GloVe embeddings} \cite{pennington2014glove} on the labels} and NSFW content flagging information from the pre-trained \texttt{NSFW-MobileNet-v2} model \cite{GantMann83:NSFW} to help perform a guided search of misogynistic consent-violating transgressions. This resulted in discovery of five dozen plus images\footnote{Listed in \texttt{df\_hand\_survey.csv}} across four categories: \textit{beach-voyeur-photography, exposed-private-parts, verifiably pornographic and upskirt} in the following classes: \textit{445-Bikini, 638 -maillot, 639-tank suit, 655-miniskirt} and \textit{459-brassiere} (see Figure \ref{fig:agn_card}).  
 Lastly, we harnessed literature from areas spanning from dog-ownership bias (\cite{hirschman1994_dog_gender_1},\cite{ramirez2006_dog_gender_2}) to engendering of musical instruments (\cite{wrape2016gender_music_gender}, \cite{bullerjahn2016masculine_music}) to generate analysis of subtle forms of \textit{human co-occurrence}-based gender bias in Figure \ref{fig:co_occur_card}.
\\Captured in Table \ref{tab:csv_census} are the details of the \textit{csv} formatted data assets curated for the community to build on. The CAG statistics are covered in  { df\_insightface\_stats.csv} and {df\_audit\_age\_gender\_dex.csv}. Similarly, we have also curated NSFW scoring ({df\_nsfw.csv}), Accuracy ({df\_acc\_classwise\_resnet50/\_NasNet\_mobile.csv}) and Semanticity ({df\_imagenet\_names\_umap.csv}) datasets as well. {df\_census\_imagenet\_61.csv} contains the 61 cumulative  paramaters for each of the 1000 classes (with their column interpretations in {df\_census\_columns\_interpretation.csv}). We have duly open-sourced these meta-datasets and 14 tutorial-styled Jupyter notebooks (spanning both ImageNet and Tiny-Images datasets) for community access\footnote{ Link: \url{https://rb.gy/zccdps}}.

\begin{table*}[]
\centering
\begin{tabular}{|l|l|}
\hline
Metric                  & Models used \\ \hline
Count, Age and Gender   & \texttt{DEX} \cite{rothe2018deep_dex}, \texttt{InsightFace} \cite{insight-face}, \texttt{RetinaFace} \cite{deng2019retinaface}, \texttt{ArcFace} \cite{deng2018arcface} \\ \hline
NSFW-scoring            & \texttt{NSFW-MobileNet-V2-224} \cite{GantMann83:NSFW} \\ \hline
Semanticity             & \texttt{Glove} \cite{pennington2014glove}, \texttt{UMAP} \cite{mcinnes2018umap} \\ \hline
Classification Accuracy & \texttt{Resent-50} \cite{he2016deep:ResNet}, \texttt{NasNet-mobile} \cite{zoph2018:NasNet} \\ \hline
\end{tabular}
\caption{Metrics considered and pre-trained models used}
\label{tab:models}
\end{table*}
\begin{table*}
\centering
\captionsetup{justification=centering}
\begin{tabular}{rrrrrrrrr}
\toprule
 $N_{train-O}^{(dex)}$ &  $n_{train-O}^{(if)}$ &  $n_{val-O}^{(if)}$ &  $N_{train-O}^{(if)}$ &  $N_{val-O}^{(if)}$ & $N_{train-W}^{(if)}$ &  $N_{train-M}^{(if)}$ & $N_{val-W}^{(if)}$ & $N_{val-M}^{(if)}$\\
\midrule
                132,201 &           80,340 &          3,096 &                 97,678 &                3,392 &          26,195 &        71,439 &          645 &       2,307 \\
\bottomrule
\end{tabular}
\caption{ \textit{Humans of the imagenet dataset}: How many? \\ Key: $\left\{ {n/N} \right\}_{\{ train/val\}  - \{ O/W/M\} }^{(\left\{ {dex/if} \right\})}$.(O:Overall,W:Women,M: Men) }
\label{tab:summary}
\end{table*}
\begin{table*}[]
\begin{tabular*}{\textwidth}{c @{\extracolsep{\fill}} |c|c|c|c|c|c|}
\toprule
 class\_number &label &  mean\_gender\_audit &  mean\_age\_audit &  mean\_nsfw\_train \\
\midrule
  445 &        bikini, two-piece &           0.18 &       24.89 &         0.859 \\
  638 &                  maillot &           0.18 &       25.91 &         0.802 \\
  639 &        maillot, tank suit &           0.18 &       26.67 &         0.769 \\
  655 &           miniskirt, mini &           0.19 &       29.95 &         0.62 \\
  459 &    brassiere, bra, bandeau &           0.16 &       25.03 &         0.61 \\
\bottomrule
\end{tabular*}
\caption{Table of the 5 classes for further investigation that emerged from the NSFW analysis}
\label{tab:agn}
\end{table*}
\section{Conclusion and discussion}
We have sought to draw the attention of the machine learning community towards the societal and ethical implications of large scale datasets, such as the problem of non-consensual images and the oft-hidden problems of categorizing people. ImageNet has been championed as one of the most incredible breakthroughs in computer vision, and AI in general. We indeed celebrate ImageNet's achievement and recognize the creators' efforts to grapple with some ethical questions. Nonetheless, ImageNet as well as other large image datasets remain troublesome. 
In hindsight, perhaps the ideal time to have raised ethical concerns regarding LSVD curation would have been in 1966 at the birth of \textit{The Summer Vision Project} \cite{papert1966summer}. The right time after that was when the creators of ImageNet embarked on the project to ``map out the entire world of objects''. Nonetheless, these are crucial conversations that the computer vision community needs to engage with \textbf{now} given the rapid democratization of imaging scraping tools (\cite{bingscra68:online_pip1,ImageScr23:online_pip2,imagebot60:online_pip3}) and \textit{dataset-zoos} (\cite{Google_Dataset_search:online,DS_TensorFl10:online,DS_torchvis46:online}). The continued silence will only serve to cause more harm than good in the future. 
In this regard, we have outlined a few solutions, including \textit{audit cards}, that can be considered to ameliorate some of the concerns raised. We have also curated meta-datasets and open-sourced the code to carry out quantitative auditing using the ILSVRC2012 dataset as a template. However, we posit that the deeper problems are rooted in the wider structural traditions, incentives, and discourse of a field that treats ethical issues as an afterthought. A field where \textit{in the wild} is often a euphemism for \textit{without consent}. We are up against a system that has veritably mastered \textit{ethics shopping}, \textit{ethics bluewashing}, \textit{ethics lobbying}, \textit{ethics dumping}, and \textit{ethics shirking} \cite{floridi2019translating}. 

Within such an ingrained tradition, even the most thoughtful scholar can find it challenging to pursue work outside the frame of the ``tradition''. Subsequently, radical ethics that challenge deeply ingrained traditions need to be incentivised and rewarded in order to bring about a shift in culture that centres justice and the welfare of disproportionately impacted communities. We urge the machine learning community to pay close attention to the direct and indirect impact of our work on society, especially on vulnerable groups. Awareness of historical antecedents, contextual, and political dimensions of current work is imperative is this regard. We hope this work contributes in raising awareness regarding the need to cultivate a justice centred practice and motivates the constitution of IRBs for large scale dataset curation processes.

\section{Acknowledgements}

This work was supported, in part, by Science Foundation Ireland grant 13/RC/2094 and co-funded under the European Regional Development Fund through the Southern \& Eastern Regional Operational Programme  to Lero - the Irish Software Research Centre (www.lero.ie). 

The authors would like to thank Alex Hanna, Andrea E. Martin, Anthony Ventresque, Elayne Ruane, John Whaley, Mariya Vasileva, Nicolas Le Roux, Olivia Guest, Os Keyes, Reubs J. Walsh, Sang Han, and Thomas Laurent for their useful feedback on an earlier version of this manuscript.

\begin{appendices}
\section{Risk of privacy loss via reverse search engines}
\label{appendix:privacy_loss}
\begin{figure}[ht!]
 \centering
 \includegraphics[width=0.5\textwidth]{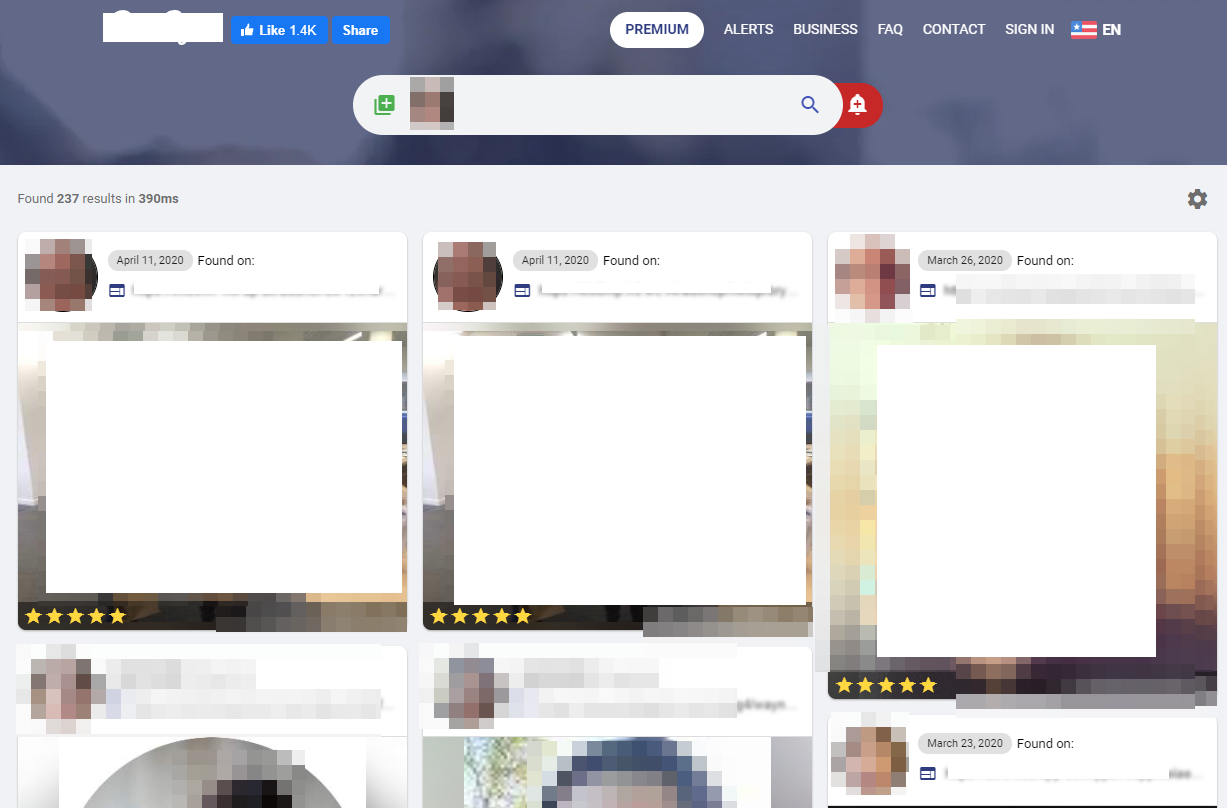}
 \caption{Snapshot of a popular reverse image search website}
 \label{fig:reverse_image}
\end{figure}

As covered in the main paper, reverse image search engines that facilitate face search such as \cite{Facesear28:online} have gotten remarkably and worryingly efficient in the past year. For a small fee, anyone can use their portal or their API to run an automated process and uncover the ``real-world'' identities of the \textit{humans of ImageNet} dataset. While all genders in the imagenet dataset are under this risk, there is asymmetric risk here as the high NSFW classes such as \texttt{bra, bikini and maillot} are often the ones with higher female-to-men ratio  (See Figure \ref{fig:nsfw_scatter}). Figure \ref{fig:reverse_image} showcases a snapshot image of one such reverse image search portal to demonstrate how easy it is for anyone to access their GUI and uncover ``real world'' identities of people which can lead to catastrophic downstream risks such as blackmailing and other forms on online abuse. 

\section{Quantitative auditing}
\label{appendix:Quantitative_auditing}
\begin{figure*}[ht!]
 \centering
 \includegraphics[width=0.95\textwidth]{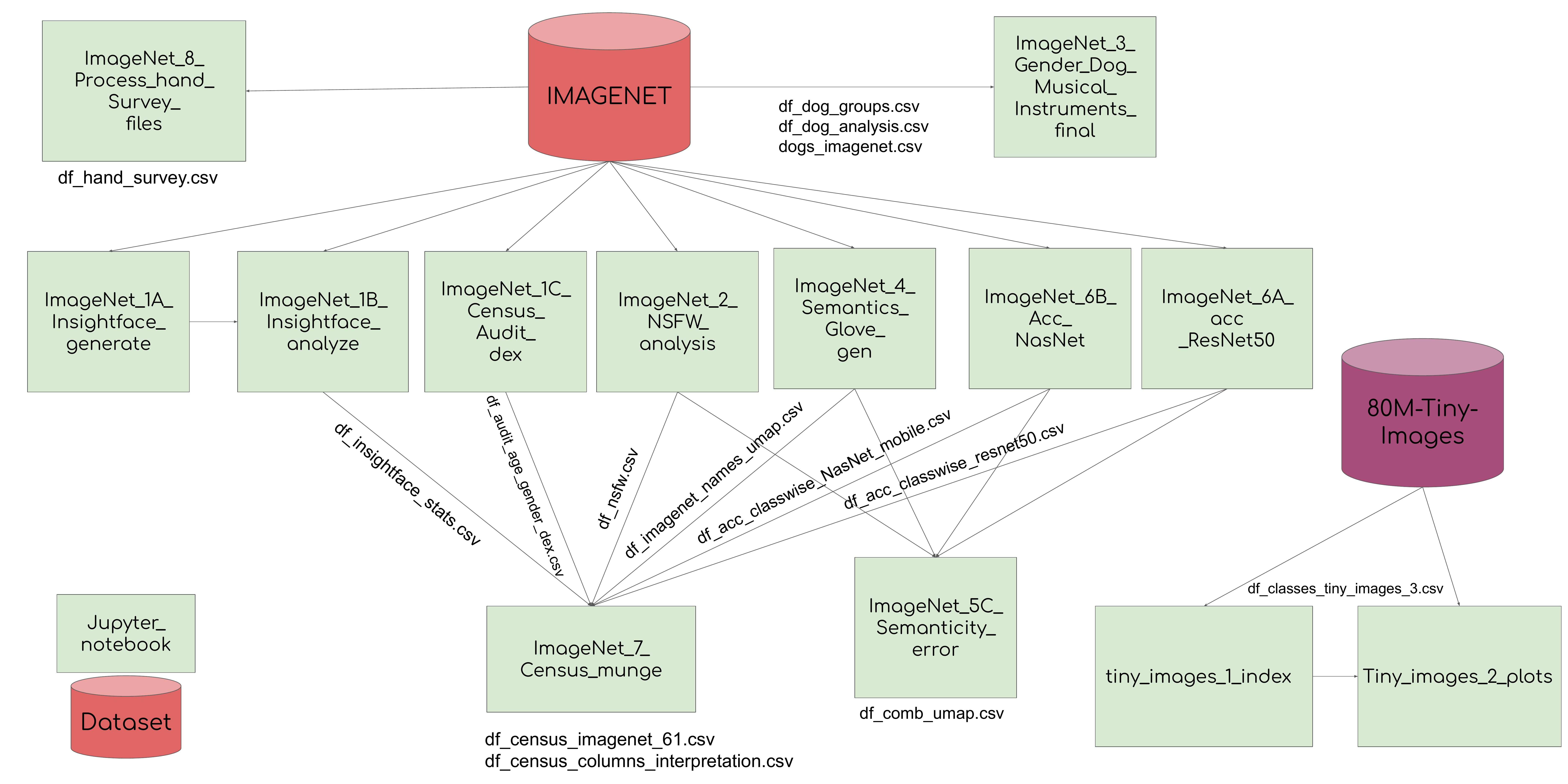}
 \caption{Visualization of all the notebooks and dataset assets curated during the quantitative analysis}
 \label{fig:assets_map}
\end{figure*}

In this section, we cover the details of performing the quantitative analysis on the ImageNet dataset including the following metrics: Person CAG (Count -Age - Gender) , NSFW scoring of the images, Semanticity  and classification accuracy. The pre-trained models used in this endeavor are covered in Table \ref{tab:models}.
All of these analyses and the generated meta-datasets have been open sourced at \url{https://rb.gy/zccdps}. 
Figure \ref{fig:assets_map} covers the details of all the jupyter notebooks authored to generate the datasets covered in Table \ref{tab:csv_census}.
\subsection{Count, Age and Gender}

\begin{figure}[!h]
  \centering
  \includegraphics[width=0.5\textwidth]{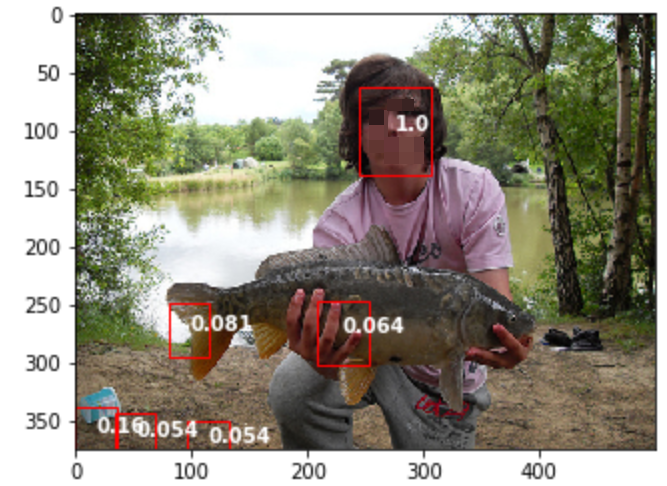}
  \caption{An example image with the output bounding boxes and the confidence scores of the humans detected in the image by the \texttt{DEX} model(\cite{rothe2018deep_dex})}
  \label{fig:dex_errors}
\end{figure}

In order to perform a human-centric census covering metrics such as count, age, and gender, we used the  \texttt{InsightFace} toolkit for face analysis \cite{insight-face}, that provided implementations of:  \texttt{ArcFace} for deep face recognition \cite{deng2019retinaface} and \texttt{Retina-Face} for face localisation (bounding-box generation) \cite{deng2019retinaface}. We then combined the results of these models with the results obtained from \cite{dulhanty2019auditing} that used the \texttt{DEX} \cite{rothe2018deep_dex} model. The results are as shown in Table \ref{tab:summary} that captures the summary statistics for the \texttt{ILSVRC2012} dataset. In this table, the lower case $n$ denotes the number of \textit{images with persons} identified in them whereas $N$ indicates the \textit{number of persons}\footnote{The difference is simply on account of more than one person being identified by the model in a given image}. The superscript indicates the algorithm used (\texttt{DEX} or \texttt{InsightFace} (if) ) whereas the subscript has two fields: The train or validation subset indicator and the census gender-category. For example, $n_{val-O}^{(if)}=3096$ implies that there were 3096 images in the ImageNet validation set (out of $50000$) where the \texttt{InsightFace} models were able to detect a person's face.

As shown, the \texttt{InsightFace} model identified 101,070 persons across 83,436 images (including the train and validation subsets) which puts the prevalence rate of persons whose presence in the dataset exists sans explicit consent to be around $7.6\%$ which is less aggressive compared to the $10.3\%$ predicted by the DEX model (that focussed on the training subset), which has a higher identification false positive rate. An example of this can be seen in Fig \ref{fig:dex_errors} which showcases an example image with the bounding boxes of the detected persons in the image.

\begin{figure}[!ht]
  \centering
  \includegraphics[width=0.5\textwidth]{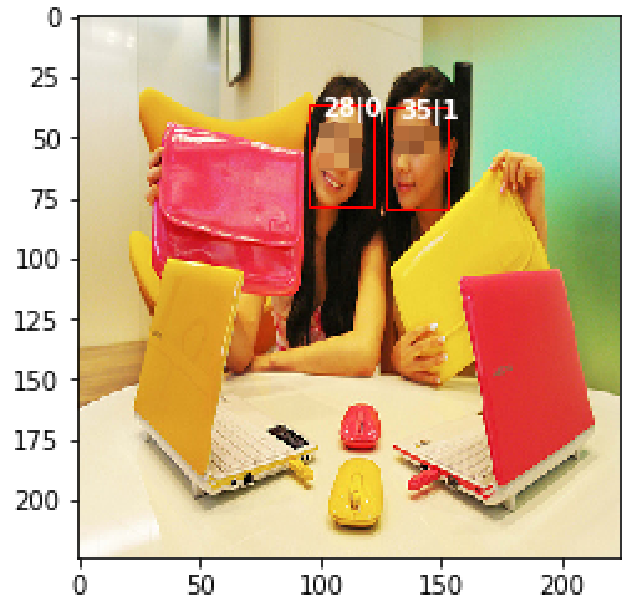}
  \caption{An example image with the output bounding boxes and the estimated ages/ (binarized) genders of the persons detected in the image by the \texttt{InsightFace} model. (Here 0: female and 1: Male)}
  \label{fig:age_errors}
\end{figure}

Much akin to \cite{dulhanty2019auditing}, we found a strong bias towards (relatively older) male presence (73,746 with a mean age of 33.24 compared to 26,840 with a mean age of 25.58). At this juncture, we would like to reemphasize that these \textit{high accuracy} pre-trained models can indeed be highly error prone conditioned on the ethnicity of the person, as analyzed in  \cite{dulhanty2019auditing,buolamwini2018gender} and we would like to invite the community to re-audit these images with better and more ethically responsible tools (See Fig \ref{fig:age_errors} for example of errors we could spot during the inference stage).

\begin{figure*}[ht!]
\centering
\subfloat[Class-wise estimates of number of humans in the images]{
  \includegraphics[width=0.95\textwidth]{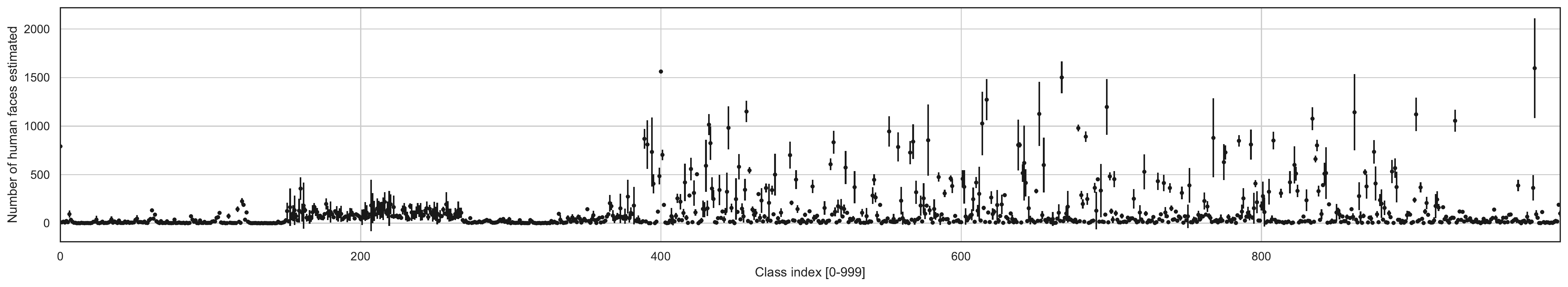}
  \label{fig:dex_if_a}
}
\hspace{0mm}
\subfloat[Scatter-plots with correlations covering the cardinality, age and gender estimates]{
  \includegraphics[width=0.95\textwidth]{plots/census/1b_dex_if_comparisons.pdf}
  \label{fig:dex_if_b}
}
\caption{Juxtaposing the results from the \texttt{DEX} and the \texttt{InsightFace} models}
\label{fig:dex_if}
\end{figure*}

Figure \ref{fig:dex_if_a}, presents the class-wise estimates of the number of persons in the dataset using the DEX and the InsightFace models. In Figure \ref{fig:dex_if_b}, we capture the variation in the estimates of count, gender and age of the \texttt{DEX} and the \texttt{InsightFace} models. 

Before delving in to the discussions of the results obtained, we define the parameters that were measured. 
To begin, we denote $\phi_i$ to be the binary \textit{face-present} indicator variable(
$
{\phi _i} = 
     \begin{cases}
       \text{1} &\text{if face present }\\
       \text{0} &\text{otherwise} \\
     \end{cases}
$) with regards to the image indexed $i$, 
$(A)$ (in the superscripts) to be the algorithm used ($A \in \{\texttt{DEX,INSIGHTFACE}\}$), and $N_c$  to be the number of images in the class $c$. Now, we define the class-level mean person count ($\eta _c^{(A)}$), mean-gender-skewness score ($\xi _c^{(A)}$) and mean-age ($\alpha_c^{(A)}$) to be, 

\begin{align*} 
\eta _c^{(A)} & = \frac{1}{{{N_c}}}\sum\limits_{i = 1}^{{N_c}} {{\phi _i}}\\
\alpha _c^{(A)} & = \frac{1}{{{N_c}}}\sum\limits_{i = 1}^{{N_c}} {{\phi _i}a_i^{(A)}}\\
\xi _c^{(A)} & = \frac{1}{{{N_c}}}{\sum\limits_{i = 1}^{{N_c}} {{\phi _i}\left( {\frac{{g_i^{(A)} - \mu _c^{(A)}}}{{\sigma _c^{(A)}}}} \right)} ^3}
\end{align*} 
Here, $a_i^{(A)}$ is the age-estimate of the person generated by algorithm $(A)$ in the $i^{th}$ image and ${\mu _c^{(A)}}$ and ${\sigma_c^{(A)}}$ represent the mean and standard-deviation of the gender-estimate of the images belonging to class $c$ and estimated by algorithm $(A)$ respectively.\\

With regards to the first scatter-plot in  Figure \ref{fig:dex_if}(b), we observe that the estimated class-wise counts of persons ($\eta _c^{(A)}$) detected by the \texttt{DEX} and \texttt{InsightFace} models in the images were in strong agreement ($Pearson-r=0.973(0.0)$) which helps to further establish the global person prevalence rate in the images to be in the order of $7.6-10.3\%$. These scatter-plots constitute Figure \ref{fig:cag_card} of the dataset audit card (Figure \ref{fig:audit_card}).
\\Now, we would like to draw the attention of the reader towards the weaker correlation ($Pearson-r=0.723(0.0)$) when it came to gender-skewness ($\xi _c^{(A)}$) and the mean age-estimates ($\alpha _c^{(A)};\ Pearson-r=0.567(0.0)$) scatter-plots in Figure \ref{fig:dex_if}(b). Given that the algorithms used are state-of-the-art with regards to the datasets they have been trained on (see \cite{rothe2018deep_dex} and \cite{insight-face}), the high disagreement on a ``neutral'' dataset like ImageNet exposes the frailties of these algorithmic pipelines upon experiencing population shifts in the test dataset. This, we believe, lends further credence to the studies that have demonstrated poor reliability of these so-termed accurate models upon change of the underlying demographics (see \cite{dulhanty2019auditing} and \cite{buolamwini2018gender}) and further supports the need to move away from gender classification on account of not just the inherent moral and ethical repugnance of the task itself but also on its lack of merit for \textit{scientific validity} \cite{ainow_1}.
\subsection{NSFW scoring aided misogynistic imagery hand-labeling}
Previous journalistic efforts (see \cite{Insideth82:online_register}) had revealed the presence of strongly misogynistic content in the ImageNet dataset, specifically in the categories of \texttt{beach-voyeur-photography, upskirt images, verifiably pornographic and exposed private-parts}. These specific four categories have been well researched in digital criminology and intersectional feminism (see \cite{henry2017not,mcglynn2017beyond,powell2018image,powell2010configuring}) and have formed the backbone of several legislations worldwide (see \cite{mcglynn2017more_law1},\cite{gillespie2019tackling_law2}). In order to help generate a hand labelled dataset of these images amongst more than 1.3 million images, we used a hybrid human-in-the-loop approach where we first formed a smaller subset of images from image classes filtered using a model-annotated NSFW-average score as a proxy. For this, we used the \texttt{NSFW-Mobilenet-v2} model \cite{GantMann83:NSFW} which is an image-classification model with the output classes being \texttt{[drawings, hentai, neutral, porn, sexy]}. We defined the NSFW score of an image by summing up the softmax values of the \texttt{[hentai, porn, sexy]} subset of classes and estimated the mean-NSFW score of all of the images of a class to obtain the results portrayed in Figure \ref{fig:agn}. On the left hand side of Figure \ref{fig:agn}, we see the scatter-plot of the mean-NSFW scores plotted against the mean-gender scores (obtained from the DEX model estimates) for the 1000 imagenet classes. We then found five natural clusters upon using the \textit{Affinity Propagation} algorithm \cite{frey2007clustering_ap}. Given the \texttt{0:FEMALE|1:MALE} gender assignments in the model we used (see \cite{dulhanty2019auditing}), classes with lower mean-gender scores allude towards a \textit{women-majority class}). The specific details of the highlighted cluster in the scatter-plot in Figure \ref{fig:agn} are displayed in Table \ref{tab:agn}. Further introducing the age dimension (by way of utilising the mean-age metric for each class), we see in the right hand side of Figure \ref{fig:agn}, that the classes with the highest NSFW scores were those where the dominating demographic was that of young women. With this shortlisting methodology, we were left with approximately 7000 images which were then hand labelled by a team of five volunteers (three male, two female, all aged between 23-45) to curate a list of $61$ images where there was complete agreement over the 4 class assignment. We have open-sourced the hand-curated list (see Table \ref{tab:hand_survey}), and the summary results are as showcased in Figure \ref{fig:hand_labelled}. In sub-figure Figure \ref{fig:hand_labelled_grid}, we see the cross-tabulated class-wise counts of the four categories of images\footnote{ This constitutes Figure \ref{fig:agn_card}( in the data audit card)} across the imagenet classes and in Figure \ref{fig:hand_labelled_histo}, we present the histogram-plots of these 61 hand-labelled images across the imagenet classes. As seen, the \texttt{bikini, two-piece} class with a mean NSFW score of $0.859$ was the \textit{main} image class with 24 confirmed \texttt{beach-voyeur} pictures. 

Here, we would like to strongly reemphasise that we are disseminating this list as a community resource so as to facilitate further scholarly engagement and also, if need be, to allow scholars in countries where incriminating laws (see \cite{durham2015opposing}) may exist, to deal with in the appropriate topical way deemed fit. We certainly admit to the primacy of \textbf{context} in which the \textit{objectionable} content appears. For example, the image \texttt{n03617480\_6206.jpeg} in the class \texttt{n03617480 - kimono} that contained genital exposure, turned out to be a photographic bondage art piece shot by Nobuyoshi Araki\cite{ozaki2008shashinjinsei} that straddles the fine line between \textit{scopophilic eroticism} and pornography. But, as explored in \cite{durham2015opposing}, the mere possession of a digital copy of this picture would be punishable by law in other nations and we believe that these factors have to be considered contextually while disseminating a large scale image dataset and should be detailed as caveats in the dissemination document.

\subsubsection{NSFW and semanticity of classes}
\label{appendix:nsfw_semanticity}
\begin{figure*}[ht!]
  \centering
  \includegraphics[width=0.95\textwidth]{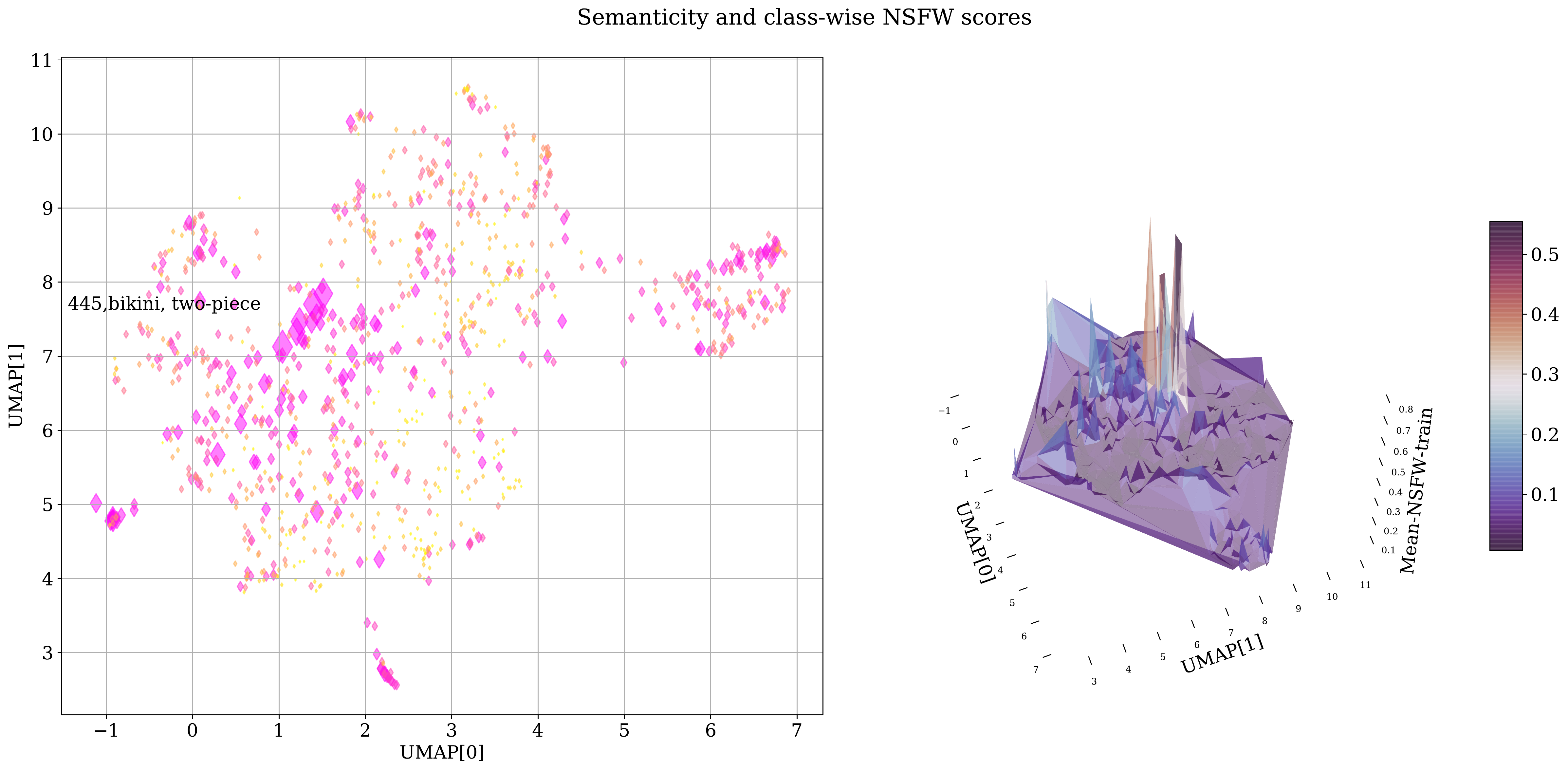}
  \caption{Figure showcasing the relationship between the semanticity of classes and the class-wise mean NSFW scores}
  \label{fig:nsfw_scatter}
\end{figure*}

\begin{figure*}[ht!]
  \centering
  \includegraphics[width=0.975\textwidth]{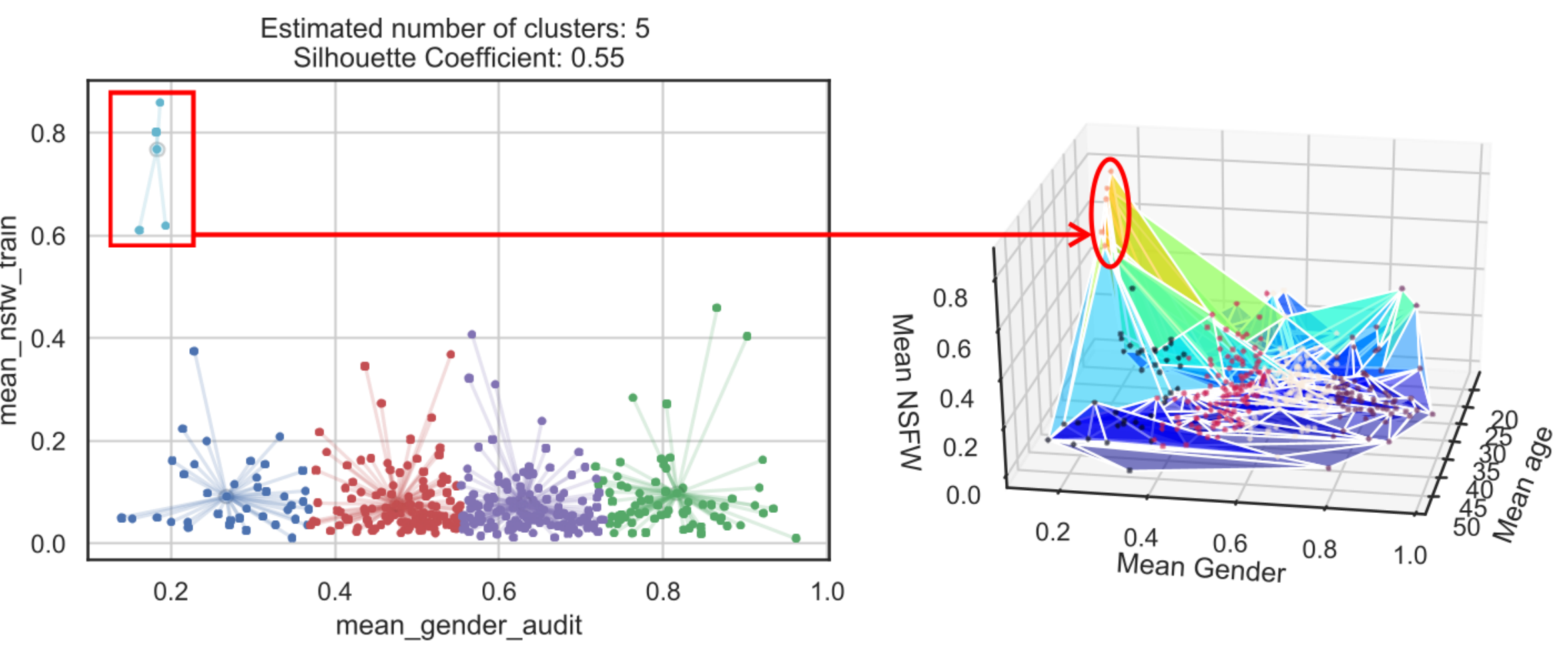}
  \caption{Class-wise cross-categorical scatter-plots across the age, gender and NSFW score estimates}
  \label{fig:agn}
\end{figure*}

\begin{figure*}[ht!]
\centering
\subfloat[Cross-tabulated grid-plot of the co-occurrence of the imagenet classes with the hand-labelled categories]{
  \includegraphics[width=0.4\textwidth]{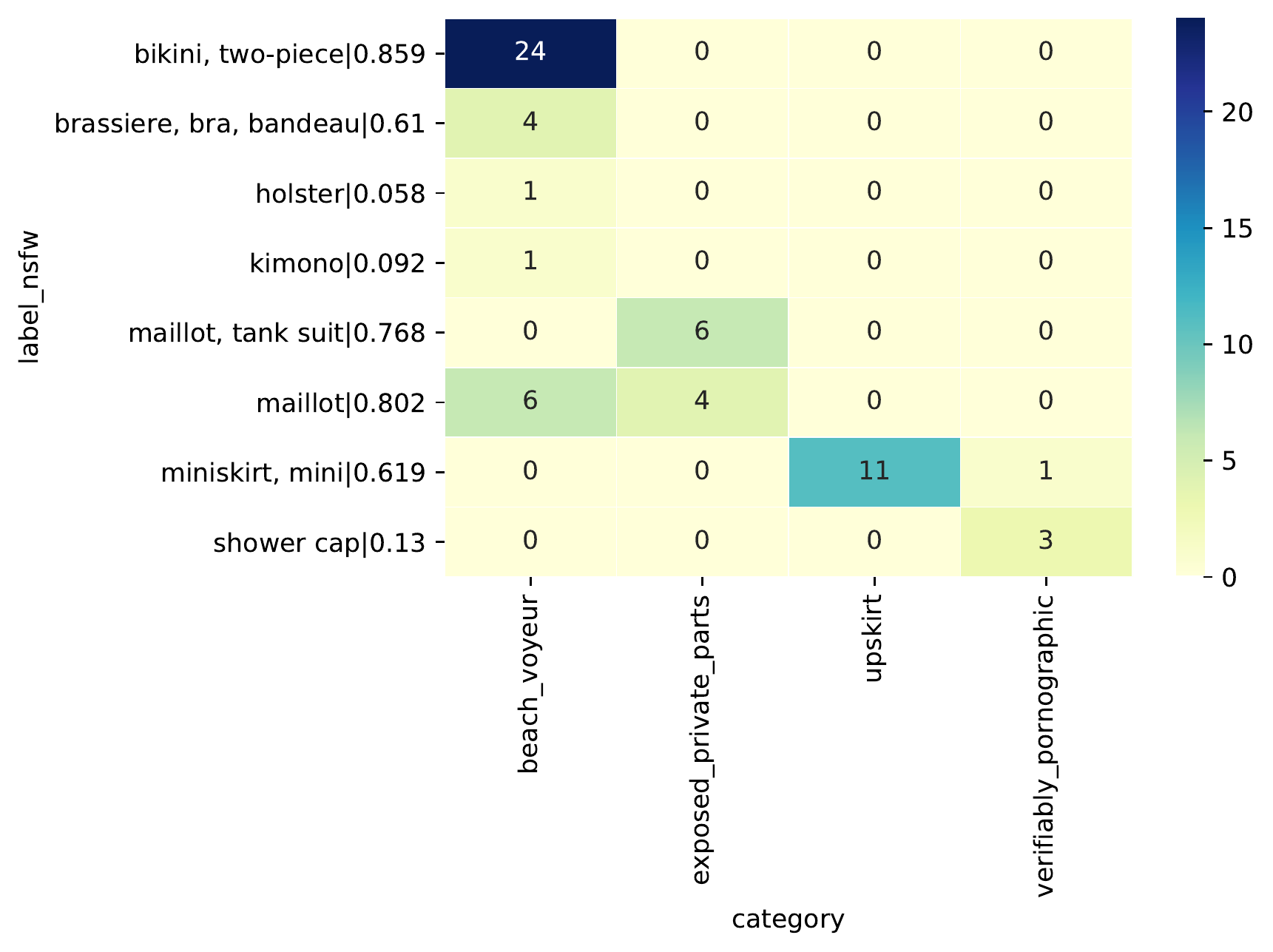}
  \label{fig:hand_labelled_grid}
}
\subfloat[Histogram-plots of the hand-labelled images]{
  \includegraphics[width=0.6\textwidth]{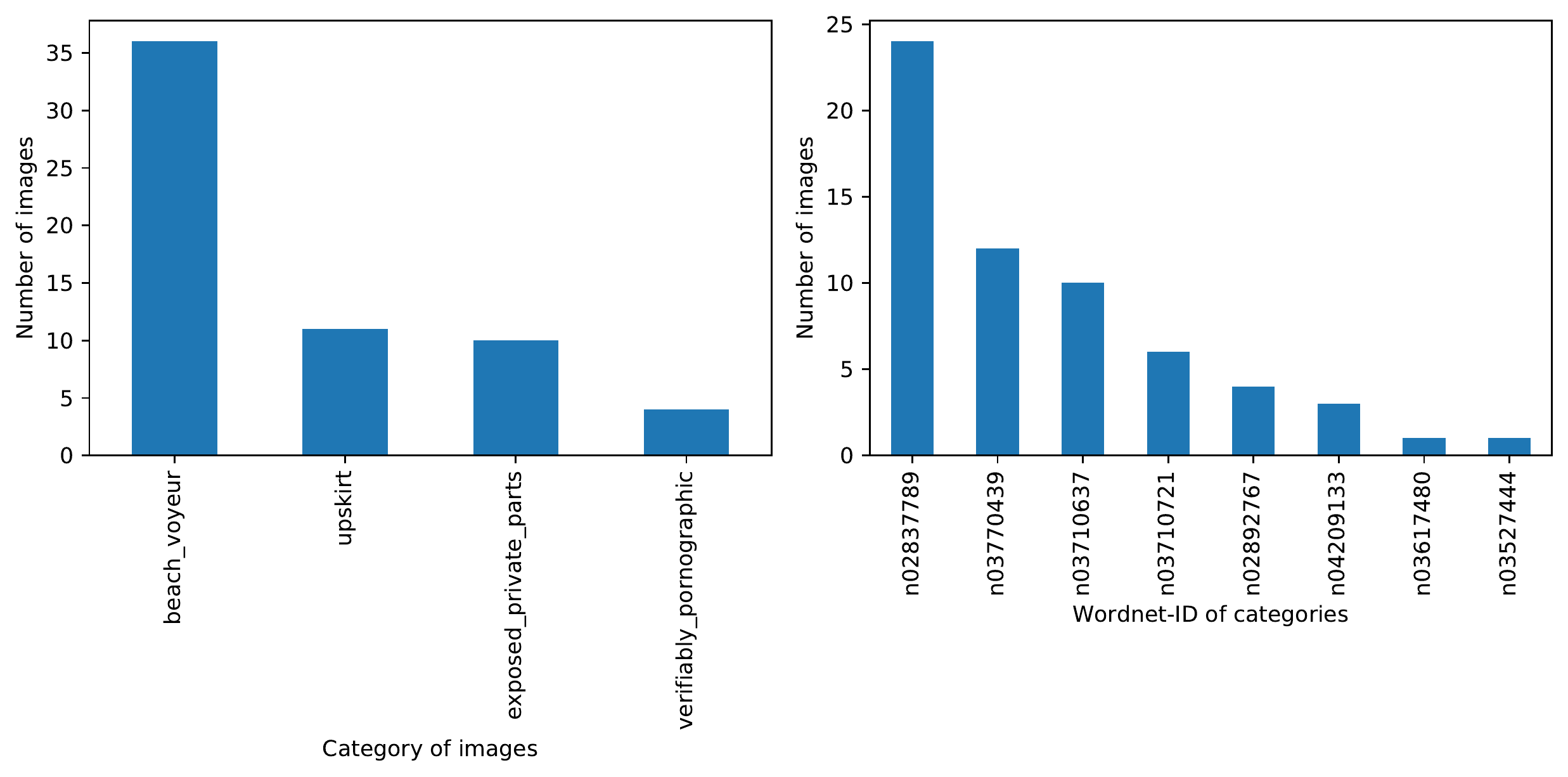}
  \label{fig:hand_labelled_histo}
}
\caption{Plots showcasing the statistics of the hand-survey across the \texttt{beach-voyeur, exposed-private-parts, upskirt, verifiably-pornographic} image categories}
\label{fig:hand_labelled}
\end{figure*}

We also analyzed the relationship between the semanticity of classes and NSFW scores. Firstly, we obtained a representative word for each of the 1000 class labels in ILSVRC2012 and used \cite{pennington2014glove} to generate dense word-vector \textit{Glove} embeddings in 300-D. Further, in order to generate the 2D/3D scatter-plots in Figure \ref{fig:nsfw_scatter}, we used the UMAP \cite{mcinnes2018umap} algorithm to perform dimensionality reduction. \texttt{df\_imagenet\_names\_umap.csv } contains the 2D UMAP embeddings of the resultant Glove vectors of the classes that are then visualized in Figure \ref{fig:nsfw_scatter} (a). In Figure \ref{fig:nsfw_scatter} (b), we see the 3D surface plot of the 2D UMAP semantic dimensions versus the NSFW scores. As seen, it is peaky in specific points of the semantic space of the label categories mapping to classes such as \texttt{brassier, bikini and maillot}.

\subsection{Dogs to musical instruments: Co-occurrence based gender biases}
\label{appendix:dogs_musical}
\begin{figure*}[ht!]
\centering
\captionsetup{justification=centering}
\subfloat[Categorized violin plot demonstrating the class-wise mean gender scores across the dog-breed image groups]{
  \includegraphics[width=0.5\textwidth]{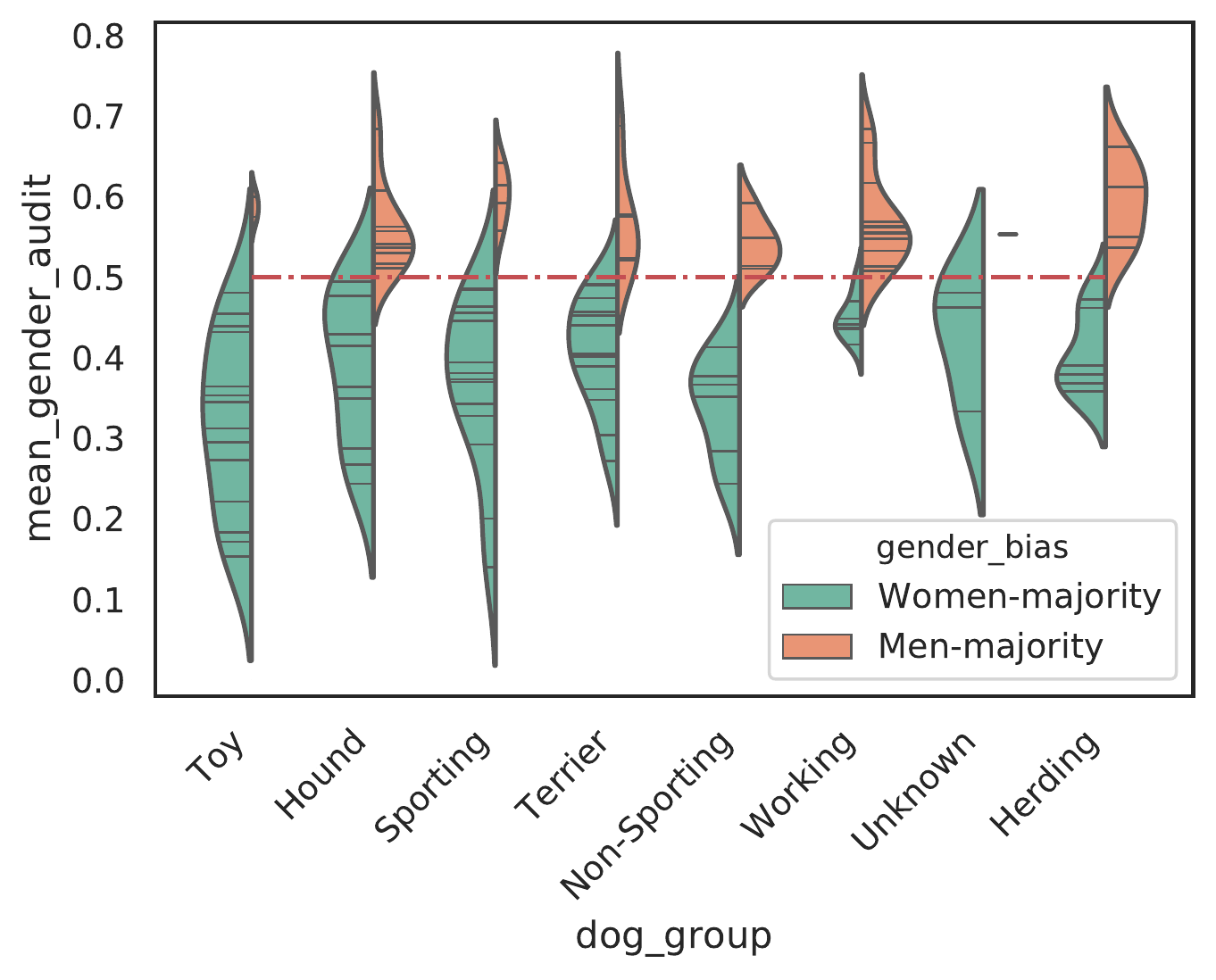}
  \label{fig:co_occur_gender_bias_dog}
}
\subfloat[Gender skewness scores across the different musical instrument image classes ]{
  \includegraphics[width=0.5\textwidth]{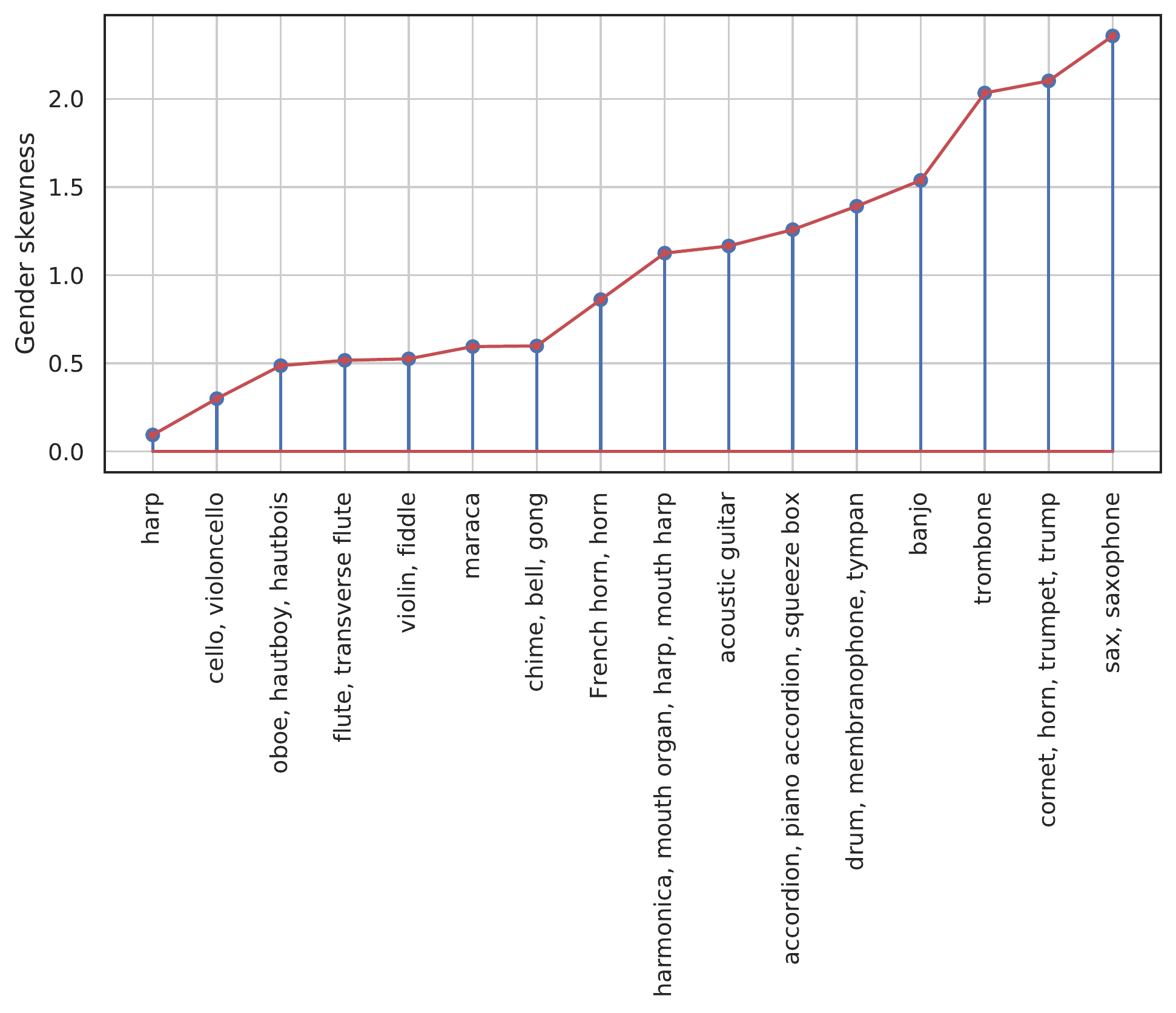}
  \label{fig:co_occur_gender_bias_instrument}
}
\caption{Plots showcasing the \textit{human co-occurrence} based gender-bias analysis}
\label{fig:co_occur_gender_bias}
\end{figure*}

Social, historical, and cultural biases prevalent in the society feed into datasets and the statistical models trained on them. In the context of Natural Language Processing (NLP), the framework of lexical co-occurrence has been harnessed to tease out these biases, especially in the context of gender biases. In \cite{tan2019assessing}, the authors analyzed occupation words stereotypically perceived as male (that they termed as \textit{M-biased} words) as well as occupation words stereotypically perceived as female (\textit{F-biased} words) in large text corpora and the ensuing downstream effects when used to generate contextual word representations in SoTA models such as such as BERT and GPT-2. Further, in \cite{Rekabsaz_Henderson_West_Hanbury_2020}, direct normalized co-occurrence associations between the word and the representative concept words were proposed as a novel corpus bias measurement method, and its efficacy was demonstrated with regards to the actual gender bias statistics of the U.S. job market and its estimates measured via the text corpora. In the context of the ImageNet dataset, we investigated if such co-occurrence biases do exist in the context of human co-occurrence in the images. Previously, in \cite{stock2018convnets_basketball}, the authors had explored the biased representation learning of an ImageNet trained model by considering the class \texttt{basketball} where images containing black persons were deemed \textit{prototypical}. Here, we tried to investigate if the gender of the person co-occurring in the background alongside the non-person class was skewed along the lines that it is purported to be in related academic work. We performed these investigations in the context of person-occurrence with regards to dog-breeds as well as musical instruments. Presented in Figure \ref{fig:co_occur_gender_bias} (a) are the conditional violin plots relating the dog-breed group of the image class of a subset of the ImageNet dataset in comparison with the with the mean gender score obtained from the \texttt{DEX} model analyses. We obtained these measurements in two phases. In the first phase, we grouped the $120$ ImageNet classes of dog-breeds in to the following 7 groups: \texttt{[Toy, Hound ,Sporting, Terrier, Non-Sporting, Working, Herding]} following the formal American Kennel Club\footnote{AKC claims that registered breeds are \textit{assigned to one of seven groups representing characteristics and functions the breeds were originally bred for.}} (AKC) groupings (see \cite{ListofBr98:AKC}). The remaining breeds not in the AKC list were placed into the \texttt{Unknown} group. Once grouped, we computed the gender-conditioned population spreads of person-concurrence using the mean-gender value of the constituent image classes obtained estimated from \cite{dulhanty2019auditing}. Prior literature (see \cite{hirschman1994_dog_gender_1,ramirez2006_dog_gender_2}) has explored the nexus between the perceived \textit{manliness} of dog groups and the ownership gender. These stereotypical associations were indeed reflected in the person co-occurrence gender distributions in Figure \ref{fig:co_occur_gender_bias_dog}, where we see that the so perceived \textit{masculine} dog groups belonging to the set \texttt{[Non-Sporting, Working, Herding]} had a stronger male-gender co-occurrence bias. \\
In a similar vein, in Figure \ref{fig:co_occur_gender_bias_instrument} we present the gender-skewness ($\xi _c^{(DEX)} = \frac{1}{{{N_c}}}{\sum\limits_{i = 1}^{{N_c}} {I[{\phi _i}]\left( {\frac{{g_i^{(DEX)} - {\mu _c}}}{{{\sigma _c}}}} \right)} ^3}$) variation amongst the co-occurring persons across the 17 imagenet musical instrument classes. Works such as \cite{delzell1992gender_music}, \cite{zervoudakes1994gender_music} and \cite{bullerjahn2016masculine_music} have explored in depth, the gender biases there exist in musical instrument selection. As stated in \cite{wrape2016gender_music_gender}, instruments such as the \texttt{cello}, \texttt{oboe}, \texttt{flute} and \texttt{violin} have been stereotypically tagged to be \textit{feminine} whereas instruments such as the \texttt{drum}, \texttt{banjo}, \texttt{trombone}, \texttt{trumpet} and the \texttt{saxophone} were the so-termed \textit{masculine} instruments in the western context. While these stereotypes represent current and historical norms, the \textit{west-centric-bias} \footnote{See \url{https://www.kaggle.com/c/inclusive-images-challenge}} of the search engine used to curate the dataset has resulted in the mirroring of these topical real-world association biases. As seen in Figure \ref{fig:co_occur_gender_bias_instrument}, \texttt{harp}, \texttt{cello}, \texttt{oboe}, \texttt{flute} and \texttt{violin} indeed had the strongest \textit{pro-women} bias where as \texttt{drum}, \texttt{banjo}, \texttt{trombone}, \texttt{trumpet} and \texttt{saxophone} were the classes with the strongest \textit{male leaning} skewness scores.

\subsection{Classes containing pictures of infants}
\label{appendix:infants}
We found this category to be particularly pertinent both under the wake of strong legislations protecting privacy of children's digital images as well as the extent of it. We found pictures of infants and children across the following 30 image classes (and possibly more): ['\texttt{bassinet}',
 '\texttt{cradle}',
 '\texttt{crib}',
 '\texttt{bib}',
 '\texttt{diaper}',
 '\texttt{bubble}',
 '\texttt{sunscreen}',
 '\texttt{plastic bag}',
 '\texttt{hamper}',
 '\texttt{seat belt}',
 '\texttt{bath towel}',
 '\texttt{mask}',
 '\texttt{bow-tie}',
 '\texttt{tub}',
 '\texttt{bucket}',
 '\texttt{umbrella}',
 '\texttt{punching bag}',
 '\texttt{maillot - tank suit}',
 '\texttt{swing}',
 '\texttt{pajama}',
 '\texttt{horizontal bar}',
 '\texttt{computer keyboard}',
 '\texttt{shoe-shop}',
 '\texttt{soccer ball}',
 '\texttt{croquet ball}',
 '\texttt{sunglasses}',
 '\texttt{ladles}',
 '\texttt{tricycle - trike - velocipede}',
 '\texttt{screwdriver}',
 '\texttt{carousel}'].
What was particularly unsettling was the prevalence of entire classes such as '\texttt{bassinet}', '\texttt{cradle}', '\texttt{crib}' and '\texttt{bib}' that had a very high density of images of infants. We believe this might have legal ramifications as well. For example, Article 8 of the European Union General Data Protection Regulation (GDPR), specifically deals with the \textit{conditions applicable to child’s consent in relation to information society services} \cite{EURLex3255:GDPR}. %
The associated \textit{Recital 38} states verbatim that \textit{Children merit specific protection with regard to their personal data, as they may be less aware of the risks, consequences and safeguards concerned and their rights in relation to the processing of personal data. Such specific protection should, in particular, apply to the use of personal data of children for the purposes of marketing or creating personality or user profiles and the collection of personal data with regard to children when using services offered directly to a child}. Further, \textbf{Article 14} of GDPR explicitly states: \textit{Information to be provided where personal data have not been obtained from the data subject}. We advocate allying with the legal community in this regard to address the concerns raised above.

\subsection{ Blood diamond effect in models trained on this dataset}
\label{appendix:blood_diamond}
 Akin to the \textit{ivory carving-illegal poaching} and \textit{diamond jewelry art-blood diamond} nexuses, we posit there is a similar moral conundrum at play here and would like to instigate a conversation amongst the neural artists in the community.
The emergence of tools such as BigGAN \cite{brock2018large_jft} and GAN-breeder \cite{GANbreeder} has ushered in an exciting new flavor of generative digital art \cite{boden2009generative}, generated using deep neural networks (see \cite{hertzmann2019aesthetics} for a survey). A cursory search on twitter\footnote{\url{https://twitter.com/hashtag/biggan?lang=en }} reveals hundreds of interesting art-works created using BigGANs. There are many detailed blog-posts\footnote{\url{https://rb.gy/pr9pwb}} on generating neural art by beginning with seed images and performing nifty experiments in the latent space of BigGANs. At the point of writing the final version of this paper, (6/26/2020, 10:34 PM PST), users on the \textit{ArtBreeder} app\footnote{\url{https://ganbreeder.app}} had \textit{generated} 64683549 images.
Further, \textit{Christie's}, the British auction house behemoth, recently hailed the selling of the neural network generated \textit{Portrait of Edmond Belamy} for an incredible $\$432,500$ as \textit{signalling the arrival of AI art on the world auction stage}\cite{christies}. 
Given the rapid growth of this field, we believe this is the right time to have a critical conversation about a particularly dark ethical consequence of using such frameworks that entail models trained on the \textit{ImageNet} dataset which has many images that are pornographic, non-consensual, voyeuristic and also entail underage nudity. We argue the use of ill-considered seed images to train the models trickles down to the final art-form in a way similar to the blood-diamond syndrome in jewelry art \cite{fishman2004diamond}. 

\begin{figure}[ht!]
\centering
\includegraphics[width=3in]{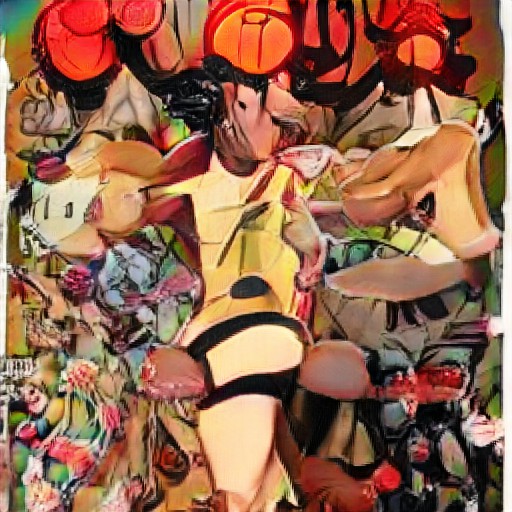}
\caption{An example neural art image generated by the authors using the \textit{ArtBreeder} app [Gene weights: Bikini: 42.35, Brassiere: 31.66, Comic Book - 84.84 ]}
\label{fig:example}
\end{figure}

\textbf{An example:} Consider the neural art image in Figure \ref{fig:example} we generated using the \textit{GanBreeder} app. On first appearance, it is not very evident as to what the constituent \textit{seed} classes are that went into the creation of this neural artwork image. When we solicited volunteers online to critique the artwork (see the collection of responses in Table \ref{tab:responses}), none had an inkling regarding a rather sinister trickle down effect at play here. As it turns out, we craftily generated this image using hand-picked specific instances of \textit{children} images emanating from what we have shown are two problematic \textit{seed} image classes: \texttt{Bikini} and \texttt{Brassiere}. More specifically, for this particular image, we set the \textit{Gene weights} to be: [\textit{Bikini}: 42.35, \textit{Brassiere}: 31.66, \textit{Comic Book} - 84.84 ]. We would like to strongly emphasize at this juncture that the problem does not emanate from a visual patriarchal mindset \cite{baran2017visual}, whereby we associate female undergarment imagery to be somehow unethical, but the root cause lies in the fact that many of the images curated into the dataset (at least with regards to the 2 above mentioned classes) were voyeuristic, pornographic, non-consensual and also entailed underage nudity.

\begin{table*}[ht]
\centering
\begin{adjustbox}{width=0.85\textwidth}
\small
\begin{tabular}{llrll}
\toprule
wordnet\_id &                    label &  mean\_nsfw\_train &                 category &                    file\_names \\
\midrule
 n02837789 &        bikini, two-piece &         0.859369 &             beach\_voyeur &          n02837789\_11383.JPEG \\
 n02837789 &        bikini, two-piece &         0.859369 &             beach\_voyeur &          n02837789\_12451.JPEG \\
 n02837789 &        bikini, two-piece &         0.859369 &             beach\_voyeur &          n02837789\_13794.JPEG \\
 n02837789 &        bikini, two-piece &         0.859369 &             beach\_voyeur &          n02837789\_14133.JPEG \\
 n02837789 &        bikini, two-piece &         0.859369 &             beach\_voyeur &          n02837789\_15158.JPEG \\
 n02837789 &        bikini, two-piece &         0.859369 &             beach\_voyeur &          n02837789\_15170.JPEG \\
 n02837789 &        bikini, two-piece &         0.859369 &             beach\_voyeur &          n02837789\_15864.JPEG \\
 n02837789 &        bikini, two-piece &         0.859369 &             beach\_voyeur &             n02837789\_17.JPEG \\
 n02837789 &        bikini, two-piece &         0.859369 &             beach\_voyeur &          n02837789\_17291.JPEG \\
 n02837789 &        bikini, two-piece &         0.859369 &             beach\_voyeur &          n02837789\_17410.JPEG \\
 n02837789 &        bikini, two-piece &         0.859369 &             beach\_voyeur &          n02837789\_18107.JPEG \\
 n02837789 &        bikini, two-piece &         0.859369 &             beach\_voyeur &          n02837789\_18124.JPEG \\
 n02837789 &        bikini, two-piece &         0.859369 &             beach\_voyeur &          n02837789\_18260.JPEG \\
 n02837789 &        bikini, two-piece &         0.859369 &             beach\_voyeur &          n02837789\_20096.JPEG \\
 n02837789 &        bikini, two-piece &         0.859369 &             beach\_voyeur &          n02837789\_22044.JPEG \\
 n02837789 &        bikini, two-piece &         0.859369 &             beach\_voyeur &            n02837789\_283.JPEG \\
 n02837789 &        bikini, two-piece &         0.859369 &             beach\_voyeur &           n02837789\_3414.JPEG \\
 n02837789 &        bikini, two-piece &         0.859369 &             beach\_voyeur &           n02837789\_3536.JPEG \\
 n02837789 &        bikini, two-piece &         0.859369 &             beach\_voyeur &              n02837789\_4.JPEG \\
 n02837789 &        bikini, two-piece &         0.859369 &             beach\_voyeur &           n02837789\_5713.JPEG \\
 n02837789 &        bikini, two-piece &         0.859369 &             beach\_voyeur &           n02837789\_9181.JPEG \\
 n02837789 &        bikini, two-piece &         0.859369 &             beach\_voyeur &           n02837789\_9859.JPEG \\
 n02837789 &        bikini, two-piece &         0.859369 &    exposed\_private\_parts &          n02837789\_17069.JPEG \\
 n02837789 &        bikini, two-piece &         0.859369 &    exposed\_private\_parts &          n02837789\_19619.JPEG \\
 n02892767 &  brassiere, bra, bandeau &         0.610233 &    exposed\_private\_parts &          n02892767\_19629.JPEG \\
 n02892767 &  brassiere, bra, bandeau &         0.610233 &    exposed\_private\_parts &           n02892767\_3235.JPEG \\
 n02892767 &  brassiere, bra, bandeau &         0.610233 &                  upskirt &          n02892767\_17717.JPEG \\
 n02892767 &  brassiere, bra, bandeau &         0.610233 &  verifiably\_pornographic &           n02892767\_5914.JPEG \\
 n03527444 &                  holster &         0.058000 &    exposed\_private\_parts &          n03527444\_12661.JPEG \\
 n03617480 &                   kimono &         0.091925 &    exposed\_private\_parts &           n03617480\_6206.JPEG \\
 n03710637 &                  maillot &         0.801976 &             beach\_voyeur &  ILSVRC2012\_val\_00021081.JPEG \\
 n03710637 &                  maillot &         0.801976 &             beach\_voyeur &          n03710637\_15836.JPEG \\
 n03710637 &                  maillot &         0.801976 &             beach\_voyeur &            n03710637\_272.JPEG \\
 n03710637 &                  maillot &         0.801976 &             beach\_voyeur &           n03710637\_3832.JPEG \\
 n03710637 &                  maillot &         0.801976 &             beach\_voyeur &           n03710637\_5095.JPEG \\
 n03710637 &                  maillot &         0.801976 &             beach\_voyeur &           n03710637\_5373.JPEG \\
 n03710637 &                  maillot &         0.801976 &             beach\_voyeur &           n03710637\_5386.JPEG \\
 n03710637 &                  maillot &         0.801976 &             beach\_voyeur &             n03710637\_66.JPEG \\
 n03710637 &                  maillot &         0.801976 &             beach\_voyeur &           n03710637\_7074.JPEG \\
 n03710637 &                  maillot &         0.801976 &    exposed\_private\_parts &           n03710637\_6756.JPEG \\
 n03710721 &       maillot, tank suit &         0.768278 &             beach\_voyeur &           n03710721\_1812.JPEG \\
 n03710721 &       maillot, tank suit &         0.768278 &             beach\_voyeur &           n03710721\_3040.JPEG \\
 n03710721 &       maillot, tank suit &         0.768278 &             beach\_voyeur &           n03710721\_3488.JPEG \\
 n03710721 &       maillot, tank suit &         0.768278 &             beach\_voyeur &           n03710721\_7542.JPEG \\
 n03710721 &       maillot, tank suit &         0.768278 &             beach\_voyeur &           n03710721\_8122.JPEG \\
 n03770439 &          miniskirt, mini &         0.619425 &                  upskirt &          n03770439\_10283.JPEG \\
 n03770439 &          miniskirt, mini &         0.619425 &                  upskirt &          n03770439\_18237.JPEG \\
 n03770439 &          miniskirt, mini &         0.619425 &                  upskirt &           n03770439\_2462.JPEG \\
 n03770439 &          miniskirt, mini &         0.619425 &                  upskirt &           n03770439\_2920.JPEG \\
 n03770439 &          miniskirt, mini &         0.619425 &                  upskirt &           n03770439\_3615.JPEG \\
 n03770439 &          miniskirt, mini &         0.619425 &                  upskirt &           n03770439\_4096.JPEG \\
 n03770439 &          miniskirt, mini &         0.619425 &                  upskirt &           n03770439\_4203.JPEG \\
 n03770439 &          miniskirt, mini &         0.619425 &                  upskirt &           n03770439\_6214.JPEG \\
 n03770439 &          miniskirt, mini &         0.619425 &                  upskirt &           n03770439\_8550.JPEG \\
 n03770439 &          miniskirt, mini &         0.619425 &                  upskirt &           n03770439\_9676.JPEG \\
 n03770439 &          miniskirt, mini &         0.619425 &  verifiably\_pornographic &          n03770439\_12003.JPEG \\
 n03770439 &          miniskirt, mini &         0.619425 &  verifiably\_pornographic &           n03770439\_1347.JPEG \\
 n04209133 &               shower cap &         0.130216 &    exposed\_private\_parts &          n04209133\_10606.JPEG \\
 n04209133 &               shower cap &         0.130216 &    exposed\_private\_parts &            n04209133\_206.JPEG \\
 n04209133 &               shower cap &         0.130216 &    exposed\_private\_parts &            n04209133\_716.JPEG \\
\bottomrule
\end{tabular}
\end{adjustbox}
\caption{ Table containing the results of hand surveyed images}
\label{tab:hand_survey}
\end{table*}

\begin{table*}[]
\begin{tabular}{|l|l|}
\hline
Reviewer-ID                           & Review                                                                                                                                                                                     \\ \hline
A- Grad student, CMU SCS              & \begin{tabular}[c]{@{}l@{}}This one reminds me of a mix between graffiti and paper mache using \\ newspaper with color images or magazines . My attention is immediately \\ drawn to near the top of the image  which, at first glance, appears to be a \\ red halo of sorts,  but upon further consideration, looks to be long black \\ branching horns on a glowing red background. \\ My attention then went to the center top portion, where the "horns" were \\ coming from, which appeared to be the head or skull of a moose or \\ something similar. The body of the creature appears to be of human-like\\  form in a crucifix position, of sorts. The image appears more and more \\ chaotic the further down one looks.\end{tabular} \\ \hline
B- Grad student, Stanford CS          & \begin{tabular}[c]{@{}l@{}}Antisymmetric: left side is very artistic, rich in flavor and shades; \\ right is more monotonic but has more texture. \\ Reminds me of the two different sides of the brain through the anti-symmetry\end{tabular}                                                                                                                                                                                 \\ \hline
C- Data Scientist, Facebook Inc       & Futurism                                                                                                                                                                              \\ \hline
D- CS undergrad, U-Michigan           & \begin{tabular}[c]{@{}l@{}}It’s visually confusing in the sense that I couldn’t tell if I was \\ looking at a 3D object with a colorful background or a painting. \\ It’s not just abstract, but also mysteriously detailed\\  in areas to the point that I doubt that a human created these\end{tabular}                                                                                                                                                                                                                                                                                                                                                                                                                                       \\ \hline
E - Senior software engineer, Mt View & \begin{tabular}[c]{@{}l@{}}The symmetry implies a sort of intentionally. \\ I get a sense of Picasso mixed with Frieda Callo[sic] here.\end{tabular}                                                                                                                                                                                                                                                                                                                                                                                                                                                                                                                                                                                                 \\ \hline
F- Data Scientist, SF                 & \begin{tabular}[c]{@{}l@{}}Reminds me of a bee and very colorful flowers, but with some \\ nightmarish masks hidden in some places. Very tropical\end{tabular}                                                                                                                                                                                                                                                                                                                                                                                                                                                                                                                                                                                  \\ \hline
\end{tabular}
\caption{Responses received for the neural art image in Fig \ref{fig:example}}
\label{tab:responses}
\end{table*}
\subsection{Error analysis}
\begin{figure*}[ht!]
  \centering
  \includegraphics[width=1\textwidth,]{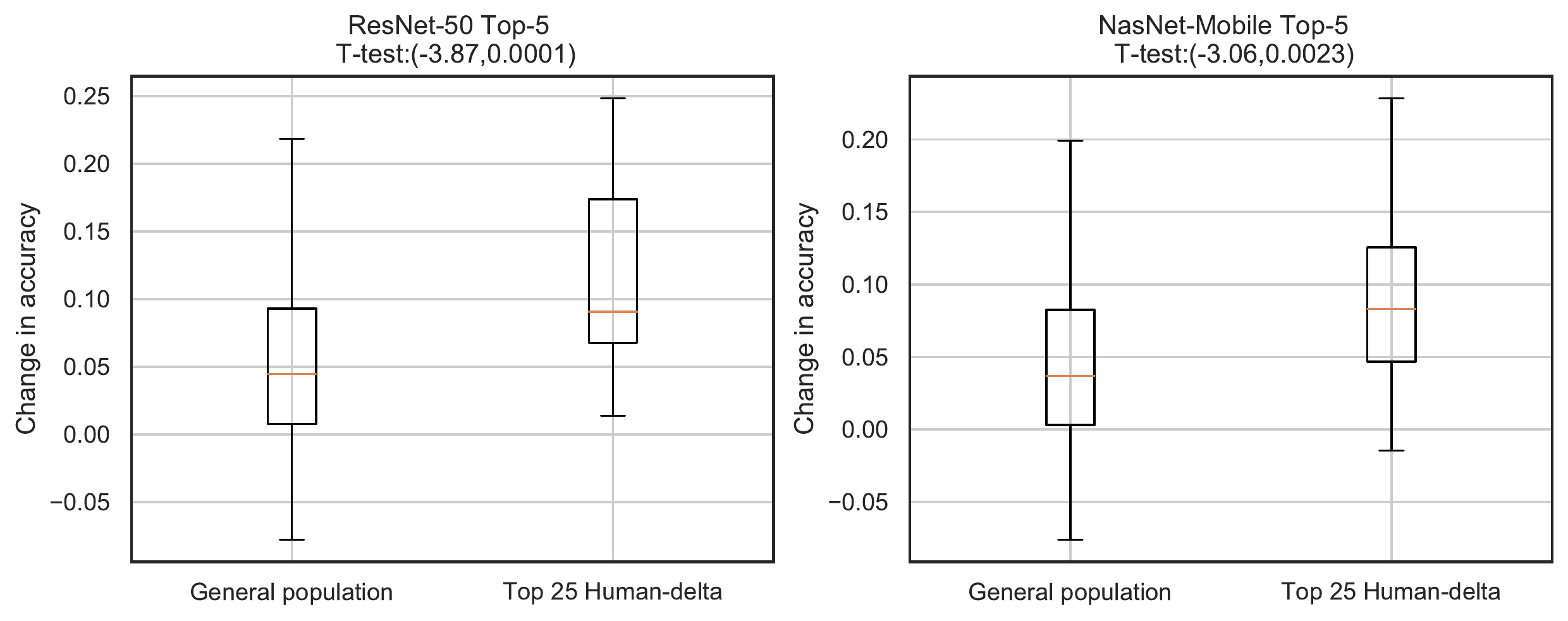}
  \caption{On accuracy variations and \textit{human delta}}
  \label{fig:acc_human_delta}
\end{figure*}

Given how besotted the computer vision community is with regards to classification accuracy metrics, we decided to indulge in \textit{devil's advocacy} by delving into the nature of variation of class-wise top-5 accuracies in those classes where humans co-occur asymmetrically between the training and validation sets. For this, we performed inference using the ResNet50 \cite{he2016deep:ResNet} and NasNet \cite{zoph2018:NasNet} models and sorted all the 1000 classes as per the $N_{train}^{persons}/N_{val}^{persons}$  ratios (termed \textit{human-delta} in the figure) and compared their accuracies with regards to the general population (amongst the 1000 classes). As gathered from Figure \ref{fig:acc_human_delta}, we saw a statistically significant drop in top-5 accuracies ($T-test \in (-3.87,-3.06)$) for the \textit{top-25 human-delta classes}, thereby motivating that even for the purveyors of \textit{scientism} fuelled pragmatism, there is motivation here to pay heed to the problem of humans in images. We would like to reemphasize that we are most certainly not advocating this to be the \textit{prima causa} for instigating a cultural change in the computer vision community, but are sharing these resources and nuances for further investigation.

\section{Broader impact statement and a wish list}
\label{appendix:impact_statement}
We embarked on this project with an aspiration to illustrate how problematic large scale image dataset curations are both in academia and industry and the need for a fundamental change. Through the course of this work, we solicited and incorporated feedback from scholars in the field who have pointed us towards three valid critiques that we would like to address first. To begin with, we solemnly acknowledge the moral paradox in our use of pre-trained gender classification models for auditing the dataset and duly address this in the previous section. Secondly, as covered in Section \ref{sec: threats} on the \textit{threat landscape}, we also considered the risks of the possible \textit{Streissand effect} with regards to deanonymization of the persons in the dataset that ultimately lead us to not dive further into the quantitative or qualitative aspects of our findings in this regard, besides conveying a specific example via email to the curator of the dataset from which the deanonymization arose. Thirdly, we would like to acknowledge the continued efforts of ImageNet curators to improve the dataset. Although there remains much work to be done, in the large scheme of things and compared to secretive and opaque datasets, the ImageNet dataset allows examinations. Having said that, curating large datasets comes with responsibility (especially given such dataset directly or indirectly impact individual lives and the social world) and all curators need to be held accountable for what they create. With these caveats firmly in tow, we now proceed to conclude with the following Wish List of the impact we hope this work may bring about.

\subsection{Proactive approach over reactive course corrections}
We aspire to see the institutions and individulas curating these large scale datasets to be proactive in establishing the primacy of ethics in the dataset curation process and not just reacting to exposes and pursing posthoc course corrections as an  afterthought. We would be well served to remind ourselves that it took the community 11 years to go from the first peer-reviewed dissemination \cite{deng2009imagenet} of the imagenet dataset to achieving the first meaningful course correction in \cite{yang2020towards_facct} whereas the \textit{number of floating-point operations required to train a classifier to AlexNet-level performance on ImageNet had decreased by a factor of 44x between 2012 and 2019} \cite{openai_moores}. This, we believe, demonstrates where the priorities lie and this is precisely where we seek to see the most impact.

\subsection{Bluewashing of AI ethics and revisiting the enterprise of Big data}
At the outset, we question if \textit{Big Data} can ever operate in a manner that caters the needs and welfares of marginalized communities - those disproportionately impacted by algorithmic injustice. Automated large scale data harvesting forays, by their very volition, tend to be \textit{BIG}, in the sense that they are inherently prone to \textbf{B}ias, are \textbf{I}mperceptive to the lessons of human condition and recorded history of vulnerable people and \textbf{G}uileful to exploit the loopholes of legal frameworks that allow siphoning off of lived experiences of disfranchised individuals who have little to no agency and recourse to contend Big Data practices. Both collective silence and empty lip service \footnote{\url{https://www.media.mit.edu/articles/beware-corporate-machinewashing-of-ai/}}, i.e. caricatured appropriations of ethical transgressions entailing \textit{ethics shopping}, \textit{ethics bluewashing}, \textit{ethics lobbying}, \textit{ethics dumping} and \textit{ethics shirking} \cite{floridi2019translating} cause harm and damage. Given that these datasets emerged from institutions such as Google, Stanford, NYU and MIT, all with a substantial number of staff researching AI ethics and policy, we cannot help but feel that this hints towards not just compartmentalization and fetishization of ethics as a \textit{hot topic} but also shrewd usage of the ethicists as agents of \textit{activism outsourcing}.

\subsection{Arresting the creative commons loot}
As covered in the main paper, we could like to see this trend of using the creative commons loophole as an excuse for circumventing the difficult terrain of informed consent. We should, as a field, aspire to treat consent in the same rigorous way as researchers and practitioners in fields such as anthropological studies or medical studies.
In this work, we have sought to draw the attention of the Machine Learning community towards the societal and ethical implications of large scale datasets, such as the problem of non-consensual images and the oft-hidden problems of categorizing people. We were inspired by the adage of \textit{Secrecy begets tyranny}\footnote{From Robert A. Heinlein's 1961 science fiction novel titled \textit{Stranger in a Strange Land \cite{heinlein2014stranger}}} and wanted to issue this as a call to the Machine Learning community to pay close attention to the direct and indirect impact of our work on society, especially on vulnerable groups. We hope this work contributes to raising awareness and adds to a continued discussion of ethics in Machine Learning, along with many other scholars that have been elucidating algorithmic bias, injustice, and harm.  
\end{appendices}

{\small
\bibliographystyle{ieee_fullname}
\bibliography{ref.bib}
}

\end{document}